\documentclass[a4paper,11pt]{article}
\pdfoutput=1 % if your are submitting a pdflatex (i.e. if you have
\usepackage{jheppub} % for details on the use of the package, please
\usepackage[T1]{fontenc} % if needed

\usepackage{amsmath,amssymb}

\usepackage{braket}

\usepackage{bm}

\usepackage{here}

\usepackage{subcaption}

\usepackage{slashed}

\title{\fontsize{20pt}{10pt}\selectfont
Entanglement Entropy and Decoupling \\[1ex]
in the Universe
}

\author[a,b]{Yuichiro Nakai,}
\author[a]{Noburo Shiba}
\author[c]{and Masaki Yamada}

\affiliation[a]{Department of Physics, Harvard University, Cambridge, MA 02138, USA}
\affiliation[b]{Department of Physics and Astronomy, Rutgers University, Piscataway, NJ 08854, USA}
\affiliation[c]{Institute of Cosmology, Department of Physics and Astronomy, Tufts University, Medford, MA 02155, USA}

\emailAdd{ynakai@physics.harvard.edu}
\emailAdd{nshiba@fas.harvard.edu}
\emailAdd{Masaki.Yamada@tufts.edu}

\abstract{In the expanding universe, two interacting fields are no longer in thermal contact when the interaction rate becomes smaller than the Hubble expansion rate. After decoupling, two subsystems are usually treated separately in accordance with equilibrium thermodynamics and the thermodynamic entropy gives a fiducial quantity conserved in each subsystem. In this paper, we discuss a correction to this paradigm from quantum entanglement of two coupled fields. The thermodynamic entropy is generalized to the entanglement entropy. We formulate a perturbation theory to derive the entanglement entropy and present Feynman rules in diagrammatic calculations. For specific models to illustrate our formulation, an interacting scalar-scalar system, quantum electrodynamics, and the Yukawa theory are considered. We calculate the entanglement entropy in these models and find a quantum correction to the thermodynamic entropy. The correction is revealed to be important in circumstances of instantaneous decoupling.}

\begin{document} 
\maketitle
\flushbottom

\section{Introduction}

\begin{figure}[!t]
\vspace{-1.5cm}
  \begin{center}
   \includegraphics[clip, width=12cm]{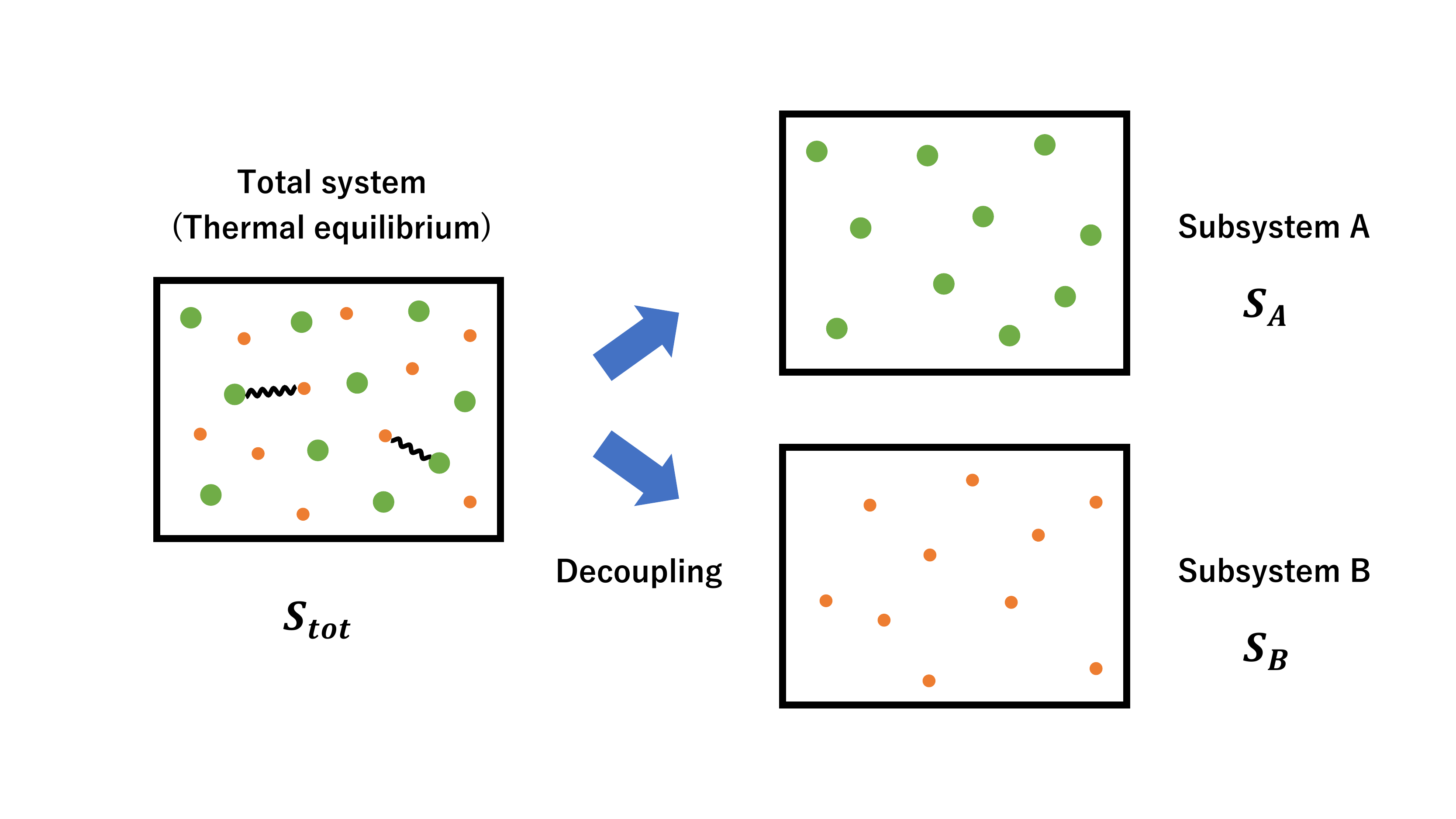}
  \end{center}
\vspace{-0.8cm}
  \caption{A schematic picture of decoupling of particles $A$ (large green dots) and $B$ (small orange dots).
The initial total system is in thermal equilibrium (particles $A$ and $B$ are interacting with each other)
and the thermodynamic entropy $S_{\rm tot}$ is a conserved quantity. 
After decoupling, it can be considered that particles $A$ and $B$ are no longer interacting with each other and
two subsystems are treated separately. In the case of instantaneous decoupling (see the main text),
the entanglement entropy $S_A$ ($S_B$)
becomes an appropriate conserved quantity for the subsystem $A$ ($B$). The entropy of the total system $S_{\rm tot}$
is not equal to the sum of $S_A$ and $S_B$ in general.}
  \label{fig:intro}
\end{figure}

The early Universe is well described by equilibrium thermodynamics
since reaction rates of particles are much faster than the Hubble expansion rate $H$ throughout most of its history
(see e.g. Ref.~\cite{Kolb:1990vq}).
The thermodynamic entropy in a comoving volume
is expected to be conserved and gives a fiducial quantity during the expansion of the Universe.
However, to reach the present state, the Universe experienced departures from thermal equilibrium.
Decoupling of two particle systems is a nonequilibrated process caused by cosmic expansion.
Figure~\ref{fig:intro} shows a schematic picture of this process.
Particles $A$ whose interaction rate with particles $B$
is smaller than the expansion rate at a temperature $T$,
\begin{equation}
\Gamma_{AB} = n_B \langle \sigma_{AB} |v| \rangle < H \sim T^2 / M_{\rm Pl} \, ,
\end{equation}
are no longer in thermal contact with particles $B$.
Here, $n_B$ is a number density of target particles $B$ and $\langle \sigma_{AB} |v| \rangle$ is a thermally averaged cross section times relative velocity.
The most famous example is neutrino decoupling with electrons at $T \sim 1 \, \rm MeV$ where
the interaction rate $\Gamma_{\nu e} \sim G_F^2 T^5$ ($G_F$ is the Fermi constant) decreases faster than the expansion rate.
Decoupled subsystems (particles), $A$ and $B$, are usually treated separately.
For the case of neutrino decoupling, the neutrino temperature drops independently from the temperature of photons (in thermal contact
with electrons) as the Universe expands.
In fact,  annihilations of electrons and positrons after the decoupling transfer their entropy into photons and
increase the photon temperature while neutrinos do not share this effect.
The thermodynamic entropy is expected to be conserved in each subsystem and
the sum of the entropy in two subsystems is assumed to be equal to the entropy in the total system before decoupling.
This picture is approximately correct when a subsystem keeps its own thermal equilibrium during the decoupling process
by some interactions inside the subsystem.
In other words, when decoupling of $A$ and $B$ proceeds much slower than
the time scale of self-interactions, thermal equilibrium in a subsystem is maintained.
Photons and electrons keep thermal equilibrium during neutrino decoupling because of their electromagnetic interactions.

In some situations, however, 
decoupling may occur instantaneously at least in an approximation, 
where the decoupling process proceeds much faster than not only the time scale of cosmic expansion $H^{-1}$
but also that of interactions in a subsystem.
We refer to this phenomenon as an {\it instantaneous decoupling} of particles $A$ and $B$. 
For instance, spontaneous breaking of a new gauge symmetry can give an instantaneous change of a cross section $\sigma_{AB}$.
Suppose that 
an interaction between $A$ and $B$ is mediated by a massless gauge boson 
before the symmetry breaking and
the interaction rate is estimated as $\Gamma_{AB} \sim \alpha^2 T$ ($\alpha = g^2 / 4\pi$, $g$ is the gauge coupling).
We assume that only the gauge interaction has a role in retaining thermal equilibrium between $A$ and $B$
and the time scale of the gauge interaction is much faster than that of self-interactions inside a subsystem.
After the symmetry breaking, the gauge boson becomes massive and the interaction rate changes into
$\Gamma_{AB} \sim \alpha^2 T^5 / M_V^4$ where $M_V$ is the gauge boson mass.
If $M_V$ is sufficiently large, two subsystems decouple immediately after the symmetry breaking.\footnote{
To realize this situation, we need a vacuum expectation value larger than the temperature at the symmetry breaking.
In the case of neutrinos, the interaction rate is still larger than the expansion rate just after the electroweak symmetry breaking.}
Second, consider particles $A$, $B$, $C$ whose initial mass relation is $M_A < M_B < M_C$.
Particles $A, B$ and also $B, C$ are interacting with each other.
Additionally, particles $B$ couple to a new scalar field.
When the scalar gets a vacuum expectation value, the mass of particles $B$ changes into $M_A < M_C < M'_B$.
If the interaction between $B$ and $C$ is strong (or particles $C$ have a large number of degrees of freedom)
and the $B$ decay into $A$ is suppressed,  
particles $B$ can decay into $C$ promptly.
Since the number density of $B$ drops, particles $A$ decouple from $B$.

When the instantaneous decoupling occurs, 
the thermodynamic entropy is no longer an appropriate fiducial quantity to 
describe 
a subsystem and should be replaced to its generalization, the entanglement entropy, 
due to quantum entanglement between the decoupled sectors. 
We shall describe decoupling of particles $A$ and $B$ in terms of density matrices in quantum mechanics.
Before the decoupling, the total system is in thermal equilibrium and characterized by
a Hamiltonian $H_{\rm tot}$.
The density matrix of the total system is then given by a grand canonical ensemble,
\begin{equation}
\rho_{\rm tot} = \frac{e^{-\beta (H_{\rm tot} - \mu_A \hat{N}_A - \mu_B \hat{N}_B)}}{Z_{\rm tot}} \, ,
\qquad Z_{\rm tot} = {\rm Tr}_{\mathcal{H}_{\rm tot}} e^{-\beta (H_{\rm tot} - \mu_A \hat{N}_A - \mu_B \hat{N}_B)} \, . \label{densitymatrixtotal}
\end{equation}
Here, $\beta = 1/T$ and the trace in the partition function $Z_{\rm tot}$ acts on the Hilbert space of the total system
$\mathcal{H}_{\rm tot}$.
$\mu_{A,B}$ and $\hat{N}_{A,B}$ are the chemical potential and the number operator of particles $A, B$ respectively.
In this system, the thermodynamic entropy is still well defined and can be expressed by the form of the so-called von Neumann entropy,
\begin{equation}
S_{\rm tot} = - {\rm Tr}_{\mathcal{H}_{\rm tot}} \left[ \rho_{\rm tot} \log \rho_{\rm tot} \right]  .
\end{equation}
The Hilbert space of the total system is decomposed into the direct product of the Hilbert spaces of the subsystems,
$\mathcal{H}_{\rm tot} = \mathcal{H}_A \otimes \mathcal{H}_B$. 
In terms of the subsystem $A$, the system $B$ can be seen as an environment 
and should be traced out at the time when the decoupling occurs.\footnote{
To be precise, this should be at the time when the last scattering occurs, as we will discuss in the later section.}
The entanglement entropy of the subsystem $A$ is defined as the von Neumann entropy
corresponding to the reduced density matrix $\rho_A$
after tracing out $\rho_{\rm tot}$ over the Hilbert space of the subsystem $B$,
\begin{equation}
S_A = - {\rm Tr}_{\mathcal{H}_A} \left[ \rho_A \log \rho_A \right] . \label{entanglemententropy}
\end{equation}
We can also consider a similar quantity for the subsystem $B$.
If the subsystem $A$ keeps thermal equilibrium during the decoupling process by self-interactions,
the quantum mechanical state evolves accordingly and the density matrix keeps the form of a grand canonical ensemble for the system $A$.
In this case, an entanglement with the system $B$ is absent 
and the entanglement entropy of the subsystem $A$ is equal to the thermodynamic entropy.
In the case of instantaneous decoupling, however, 
the quantum mechanical state does not change before and after the decoupling.
The density matrix $\rho_A$ is calculated by tracing out $\rho_{\rm tot}$ over the Hilbert space of $B$
just before the decoupling 
and cannot be written by the form of a grand canonical ensemble in general.
Then, the entanglement entropy $S_A$ is not necessarily equal to the thermodynamic entropy and also
$S_{\rm tot} \neq S_A + S_B$.
After the decoupling, 
the density matrix of the subsystem $A$ obeys a unitary evolution,
\begin{equation}
\rho_A (t_0) \rightarrow \rho_A (t) = e^{- i H_A (t - t_0)} \rho_A (t_0) \, e^{ i H_A (t - t_0)} \, ,
\end{equation}
where $t_0$ is the time of decoupling.
Under this unitary evolution, the entanglement entropy \eqref{entanglemententropy} is conserved
due to the nature of the trace. 
Therefore, the entanglement entropy is a very useful fiducial quantity to describe a subsystem after decoupling.
In fact, if particles $A$ enter thermal equilibrium due to the self-interaction
after the decoupling occurs, 
the density matrix of the system should be written as the form of a grand canonical ensemble 
and we can define the temperature of the subsystem from the conserved entanglement entropy.%
\footnote{
To define the temperature of a subsystem after decoupling,
some interactions inside the subsystem are needed to realize thermal equilibrium at a later time.
One possibility is to introduce an unbroken gauge group under which particles in the subsystem are charged.
Assuming a tiny gauge coupling, thermalization does not occur during the decoupling process.
However, since the interaction rate with this massless gauge boson is linear in $T$,
the subsystem enters its own thermal equilibrium at a low temperature.}

The entanglement entropy has been discussed extensively for the case of a density matrix
reduced to a spatial submanifold
(for reviews, see Refs.~\cite{Nielsen,Calabrese:2004eu,Ryu:2006ef,Nishioka:2009un,Casini:2009sr}) while
the case of field trace out has been paid attention to only in the context of condensed matter physics
or conformal field theories (CFTs).\footnote{
For an early attempt at zero temperature, see Ref.~\cite{Teresi:2010kt}.
}
References~\cite{Furukawa:2010nd,Chen:2013kba} have considered the entanglement entropy of coupled Tomonaga-Luttinger liquids
and Heisenberg antiferromagnets. 
References.~\cite{Xu:2011gn,Mollabashi:2014qfa,Mozaffar:2015bda} have calculated the entanglement (R\'{e}nyi) entropy of coupled CFTs.
Although Ref.~\cite{Xu:2011gn} presented the first order perturbation of a coupling between two CFTs,
a general formulation of the perturbative expansion is still lacking.
In this paper, we first formulate the perturbation theory to derive the entanglement entropy of coupled quantum fields
and present Feynman rules in the diagrammatic calculations.
Our formulation is generic and independent of the discussion of cosmology.
Next, we consider specific models such as
an interacting scalar-scalar system, quantum electrodynamics (QED) and the Yukawa theory,
and calculate the entanglement entropy.
The obtained quantum correction to the thermodynamic entropy is revealed to be important
in certain circumstances of instantaneous decoupling.
Since cosmological measurements such as the energy densities of dark matter and dark radiation are becoming more and more precise,
the quantum correction might be essential to test the viability of some cosmological models.

Here we summarize our logics and assumptions to relate the entanglement entropy 
with observed quantities in cosmology by presenting one scenario: 

\begin{itemize}

\item 
Subsystem $A$: a dark sector (e.g., dark radiation).

\item 
Subsystem $B$: the standard model sector.

\item 
The interaction between $A$ and $B$ is strong enough to maintain thermal equilibrium in the early Universe, 
but becomes negligible at a certain time (at the time of decoupling). 

\item 
The time scale of the decoupling process is assumed to be much faster than 
the time scales of self-interactions for each subsystem and the expansion of the Universe, 
so that the quantum mechanical state does not change before and after the decoupling. 
For simplicity, we may assume that self-interactions are absent (or turned off) around the time of decoupling. 

\item 
To define the temperature of the subsystem $T_i$ ($i = A, B$), 
we assume that self-interactions are turned on well after the time of decoupling. 
Since the entanglement entropy is conserved, 
we can use $S_i (T_i) = S_{i,0} + \Delta S_{i, {\rm Ent}}$, 
where $S_i (T_i) = (2 \pi^2 g_i / 45) V T_i^3$, $S_{i,0}$ is the usual thermodynamic entropy, and $\Delta S_{i, {\rm Ent}}$ is a correction coming from the entanglement effect. 
As a result, the temperature $T_i$ receives a correction $\Delta T_i$. 

\item 
The energy density of a subsystem can be calculated from the temperature, 
which has a correction from $\Delta T_i$. 
The energy density of a dark sector (e.g., dark radiation) can be indirectly measured by some cosmological observations. 

\end{itemize}

\noindent 
This scenario is an example where the entanglement entropy can give an observable effect. In particular, the subsystem $B$ can also be a dark sector. Also, particles in the subsystem $A$ may decay into some relativistic particles, which do not interact with particles in the subsystem $B$, after decoupling.
Even in this case, the entanglement entropy is conserved and may have a nonzero correction from the entanglement effect. 

The rest of the paper is organized as follows.
In section~\ref{Feynman}, we formulate the perturbation theory to derive the entanglement entropy of two coupled systems.
We present Feynman rules in the diagrammatic calculations.
In section~\ref{phi4}, we consider a scalar-scalar system and calculate the leading order correction of the entanglement entropy.
It is shown that divergence in the loop integral is renormalized correctly to give the physical result.
We then analyze the entanglement entropy in QED 
in section~\ref{QED} 
and in the Yukawa theory in section~\ref{Yukawa}. 
The quantum correction starts from two-loop level in these theories.
In section~\ref{cosmology}, we present a simple cosmological scenario of instantaneous decoupling and
discuss implications of the entanglement entropy in a subsystem after the decoupling.
In section~\ref{conclusion}, we conclude the paper and discuss possible future directions.
Some calculational details are summarized in appendices.

\section{Perturbation theory and diagrammatic techniques}\label{Feynman}

In this section, we present a functional path integral formalism of the entanglement entropy of coupled quantum fields.
In the limit of vanishing couplings between two subsystems,
their entanglement entropy reduces to the ordinary thermodynamic entropy.
New contributions appear as quantum corrections in the presence of interactions.
We formulate the perturbation theory
and present Feynman rules in the diagrammatic calculations.

\subsection{Path integral formulation of the entanglement entropy}

Let us consider interacting quantum fields $\phi_A (t, \bm{x})$ and $\phi_B (t, \bm{x})$ in $d+1$ dimensions
(we often omit to write dependence on $\bm{x}$ below).
Each field can be either a boson or a fermion.
We assume that the total system is in thermal equilibrium and 
the density matrix is given by a grand canonical ensemble of \eqref{densitymatrixtotal}.
According to the imaginary-time formalism in the finite-temperature field theory, 
we can write 
the partition function of the total system as 
a path integral for the imaginary time $\tau = it$ up to an overall normalization constant: 
\begin{equation}
\begin{split}
Z_{\rm tot}^{(\beta)}=  \int \mathcal{D} \phi_A \mathcal{D} \phi_B
\exp \left( -  \int_0^{\beta} d\tau \int d^d x \, \mathcal{L} (\phi_A, \phi_B) \right) .
\end{split}
\end{equation}
Here, the integration over the fields is constrained in such a way that
$\phi_A, \phi_B$ are (anti-)periodic in imaginary time $\tau \in (0, \beta)$ 
for bosonic (fermionic) fields as in the case of the ordinary finite-temperature field theory.
Throughout this paper, we neglect chemical potential because 
it is usually much smaller than the temperature of the Universe.

To derive the entanglement entropy of the subsystem $A$ (of the field $\phi_A$),
we consider the reduced density matrix $\rho_A$
after tracing out $\rho_{\rm tot}$ over the Hilbert space of the subsystem $B$.
Its matrix element is expressed as
\begin{equation}
\begin{split}
&\bra{\phi_A} \rho_A \ket{\phi'_A} \\
&= \frac{1}{Z_{\rm tot}^{(\beta)}} \int \mathcal{D} \phi_A \mathcal{D} \phi_B |_{\phi_A(0)=\phi'_A, \phi_A(\beta)=\phi_A}
\exp \left( - \int_0^{\beta} d\tau \int d^d x \, \mathcal{L} (\phi_A, \phi_B) \right) . \label{density matrix}
\end{split}
\end{equation}
Here, the functional form at the boundaries of $\tau$ are fixed by $\phi_A(0)=\phi'_A$ and $\phi_A(\beta)=\phi_A$.
The functional integration over the field $\phi_B$ has been performed with the appropriate boundary condition.
This density matrix cannot be written as the form of a grand canonical ensemble in general, 
so that we cannot use the thermodynamic entropy to describe this system. 
Instead, we need to use the entanglement entropy as we explain below.

\begin{figure}[!t]
\vspace{-1.5cm}
  \begin{center}
   \includegraphics[clip, width=11.5cm]{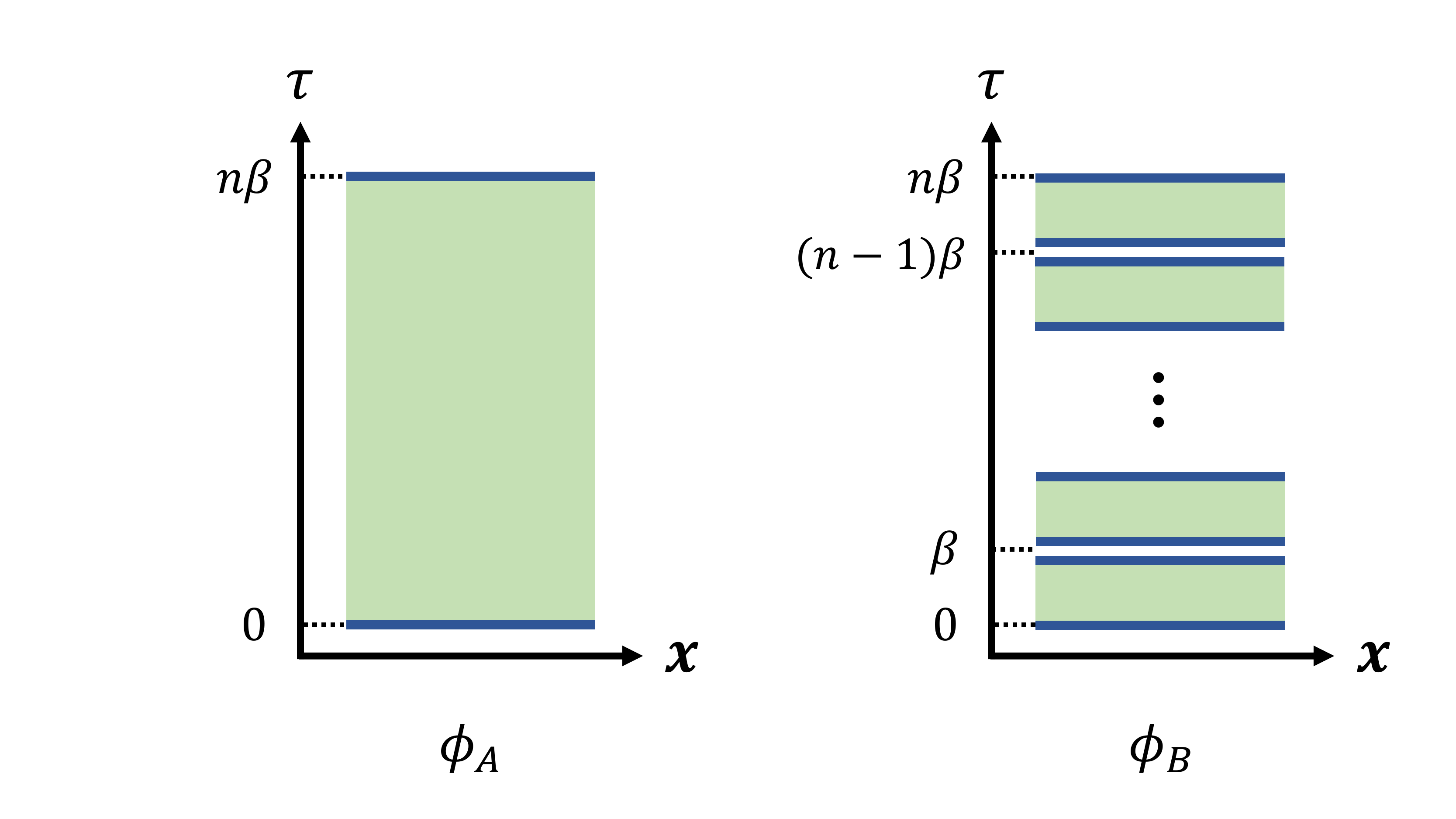}
  \end{center}
\vspace{-0.7cm}
  \caption{A schematic picture of the functional integral in $\mathrm{Tr} \rho_A^{n}$. The integration over the fields $\phi_{A,B}$
is performed in the Euclidean space $(\tau, \bm{x})$ with $\tau \in (0, n \beta)$ (green shaded region).
For the field $\phi_A$, the configuration at $\tau = n\beta$ is identified (up to sign) with the configuration at $\tau =0$.
For the field $\phi_B^{(j)}$ ($j = 1, \cdots , n$),
the configuration at $\tau = j\beta$ is identified with the configuration at $\tau =(j-1)\beta$.
In the figure, field configurations on the blue lines at the top and bottom of the shaded regions are identified.}
  \label{fig:functionalintegral}
\end{figure}

The entanglement entropy of the subsystem $A$ is defined as the von Neumann entropy
\eqref{entanglemententropy}.
Since it is not easy to evaluate the trace of $\rho_A \log \rho_A$ directly,
we first calculate 
the 
R\'{e}nyi entropy of $\phi_A$: 
\begin{equation}
S_A^{(n)}=\frac{1}{1-n} \log \left(\mathrm{Tr}  \rho_A^{n} \right) , \label{Renyi}
\end{equation}
and take the limit of $n \to 1$, which gives the entanglement entropy (${\rm lim}_{n \to 1} S_A^{(n)} = S_A$). 
This is known as the replica method.
We thus need to calculate the trace of a multiple product of the density matrix, 
which can be done 
from the expression of the reduced density matrix \eqref{density matrix}. 
Note that $\phi_B$ has a periodicity of $\beta$ because its trace is taken for each $\rho_A$, 
while $\phi_A$ has a periodicity of $n \beta$ because its trace is taken at once after the multiple product. 
Using translational invariance in imaginary time of the Lagrangian, 
we can write 
\begin{equation}
\begin{split}
\mathrm{Tr} \rho_A^{n} =\frac{\widetilde{Z}_{\rm tot}^{(n\beta)}}{ (Z_{\rm tot}^{(\beta) })^{n}} 
\equiv
\frac{1}{(Z_{\rm tot}^{(\beta)})^n} \int \mathcal{D} \phi_A \prod_j \mathcal{D}  \phi_B^{(j)} \, \exp \left( -
\sum_{j=1}^{n} \int_{(j-1)\beta}^{j\beta -\epsilon} d\tau \int d^d x
\, \mathcal{L}^{(j)} ( \phi_A , \phi_B^{(j)} )  \right) .
\label{tracerhonA}
\end{split}
\end{equation}
Here, we have introduced $n$ copies of fields $\phi_B^{(j)} \, (j=1, \cdots, n)$ each of which is defined in the interval 
$\tau \in ((j-1) \beta , \, j \beta - \epsilon)$, where 
an infinitesimal positive number $\epsilon$ is put to remind us of the boundary condition.
The field $\phi_A$ is now defined in $\tau \in (0 , n\beta)$.
The Lagrangian $\mathcal{L}^{(j)} \, (j = 1, \cdots , n)$ contains $\phi_A$ and $\phi_B^{(j)}$ 
and is defined by $\mathcal{L}^{(j)} ( \phi_A , \phi_B^{(j)} ) = 
\mathcal{L}( \phi_A , \phi_B^{(j)} )$. 
The boundary conditions of $\phi_A, \phi_B^{(j)}$ are given by
\begin{equation}
\begin{split}
&\phi_A(0)=(-1)^{F_A} \phi_A(n\beta),  \\[1.5ex]
&\phi_B^{(j)}((j-1)\beta)=(-1)^{F_B} \phi_B^{(j)}(j\beta-\epsilon) \, ,
\end{split}
\end{equation}
where $F_{A,B} = 0 \, (1)$ in the case that $\phi_{A, B}$ is a bosonic (fermionic) field.
Let us emphasize that (anti-)periodicity in imaginary time of the field $\phi_A$ is $n \beta$ while
that of the field $\phi_B^{(j)}$ is $\beta$.
It is visualized as a schematic picture in Figure~\ref{fig:functionalintegral}.
This property is essential in the following discussion.
To distinguish the difference of periodicity between $\phi_A$ and $\phi_B$,
we put a tilde in the partition function $\widetilde{Z}_{\rm tot}^{(n\beta)}$ of \eqref{tracerhonA}.

From the expression of \eqref{tracerhonA},
we can rewrite the  R\'{e}nyi entropy \eqref{Renyi} as
\begin{equation}
\begin{split}
S_A^{(n)}  &=S_{A, 0}^{(n)}  + \frac{1}{1-n} \left( \log \frac{\widetilde{Z}_{\rm tot}^{(n\beta)}}{Z_{A,0}^{(n\beta)}
 ( Z_{B,0}^{(\beta)} )^n }
-n \log \frac{Z_{\rm tot}^{(\beta)}}{Z_{A,0}^{(\beta)}  \, Z_{B,0}^{(\beta)}}  \right) ,  
\label{decomposition}
\end{split}
\end{equation}
where 
\begin{equation}
\begin{split}
Z_{\alpha, 0}^{(\beta)}  =  \int \mathcal{D} \phi_\alpha \, \exp \left( -  \int_0^{\beta} d\tau \int d^d x
\, \mathcal{L}_{\alpha,0} (\phi_\alpha) \right). 
\end{split}
\end{equation}
Here, in the partition function $Z_{\alpha, 0}^{(\beta)}$
$(\alpha = A,B)$ $\phi_\alpha$ is (anti-)periodic in imaginary time $\tau \in (0, \beta)$.
The first term of the right-hand side in Eq.~(\ref{decomposition}) gives the R\'{e}nyi entropy for the density matrix derived
from the noninteracting part of the Lagrangian $\mathcal{L}_{A, 0} (\phi_A)$: 
\begin{equation}
\begin{split}
S_{A,0}^{(n)} = \frac{1}{1-n} \log \frac{Z_{A,0}^{(n\beta)}}{(Z_{A,0}^{(\beta)})^n}  \, . \label{freeRenyi}
\end{split}
\end{equation}
In the limit of $n \rightarrow 1$, this contribution reduces to the ordinary thermodynamic entropy
of a free field.
New contributions appear from the second and third terms of \eqref{decomposition} in the presence of interactions
(note that the second and third terms include contributions to the thermodynamic entropy from
self-interactions of the field $\phi_A$).

\subsection{Zeroth-order contributions}

We here 
review 
the functional integrations in the partition functions of a free real scalar field and
a free Dirac fermion in $3+1$ dimensions.
The Euclidean Lagrangian of a neutral scalar field 
$\phi_\alpha$ ($\alpha = A, B$) is given by 
\begin{equation}
\begin{split}
\mathcal{L}_{\alpha,0} (\phi_\alpha) = \frac{1}{2} \left[ \left( \frac{\partial \phi_\alpha}{\partial \tau} \right)^2
+ \left( \nabla  \phi_\alpha \right)^2 + M_\alpha^2 \phi_\alpha^2 \right]  . \label{freescalar}
\end{split}
\end{equation}
The result of the functional integration in the partition function is
\cite{Kapusta:2006pm}
\begin{equation}
\begin{split}
\log Z_{\alpha, 0}^{(\beta)} 
= V \int \frac{d^3 p }{(2 \pi)^3}
\left\{ -\frac{1}{2} \beta \omega - \log \left( 1 - e^{- \beta \omega} \right) \right\} , \label{freescalarZ}
\end{split}
\end{equation}
where $\omega = \sqrt{\bm{p}^2 + M_\alpha^2}$ and $V$ is the volume of the system.
This is the usual form of the partition function in an ideal Bose gas with vanishing chemical potential $\mu =0$.
From this expression, we can calculate the R\'{e}nyi entropy for the free field \eqref{freeRenyi}.
Taking the limit of $n \rightarrow 1$,
we obtain the entanglement entropy as
\begin{equation}
\begin{split}
S_{\alpha,0} 
= V \int \frac{d^3 p }{(2 \pi)^3}
\left\{ \frac{\beta \omega }{e^{\beta \omega} - 1} - \log \left( 1 - e^{- \beta \omega} \right) \right\} ,
\end{split}
\end{equation}
which is equal to the thermodynamic entropy in an ideal Bose gas.
In this sense, the entanglement entropy of \eqref{decomposition} with $n \rightarrow 1$ is a generalization of
the thermodynamic entropy in the presence of interactions.

The Lagrangian of a Dirac fermion $\psi_\alpha$ in the partition function
is given by
\begin{equation}
\begin{split}
\mathcal{L}_{\alpha,0} (\psi_\alpha) = \psi_\alpha^\dagger \gamma^0 \left( \gamma^0 \frac{\partial}{\partial \tau}
 -i \bm{\gamma} \cdot \nabla +M_\alpha - \mu_\alpha \gamma^0 \right) \psi_\alpha \, .
\end{split}
\end{equation}
Here, $(\gamma^0 , \gamma^i) \, (i = 1,2,3)$ are the $4 \times 4$ Dirac matrices.
The last term in the parenthesis comes from the chemical potential.
Performing the functional integration, we obtain
\begin{equation}
\begin{split}
\log Z_{\alpha, 0}^{(\beta)} = 2V \int \frac{d^3 p }{(2 \pi)^3}
\left\{ \beta \omega + \log \left( 1 + e^{- \beta (\omega -\mu)} \right)  + \log \left( 1 + e^{- \beta (\omega +\mu)} \right)  \right\},
\label{freefermionZ}
\end{split}
\end{equation}
which is the same with the partition function in an ideal Fermi gas.
As in the case of the free scalar field,
inserting \eqref{freefermionZ} into $S_{A,0}^{(n)}$ of \eqref{freeRenyi} with the limit of $n \rightarrow 1$,
we can obtain the thermodynamic entropy in an ideal Fermi gas
with taking into account the spin degrees of freedom and the presence of antiparticles.

\subsection{Perturbative expansion and Feynman rules}

Now we 
formulate the perturbation theory of the R\'{e}nyi entropy \eqref{decomposition}
and present Feynman rules in the diagrammatic calculations.
We proceed with the discussion by considering the simplest model of a scalar-scalar system in $d+1$ dimensions, but
our procedure can be applied to any other models.
Consider two real scalar fields $\phi_A$, $\phi_B$ (the noninteracting part of their Lagrangian is given by \eqref{freescalar})
and the following interaction Lagrangian:
\begin{equation}
\begin{split}
\mathcal{L}_I = \frac{\lambda_A}{4!} \phi_A^4 + \frac{\lambda_B}{4!} \phi_B^4 + \frac{\lambda}{4} \phi_A^2 \phi_B^2 \, .
\label{scalarint}
\end{split}
\end{equation}
The first and second terms denote self-interactions and the third term is the interaction between the two fields.
This is the most general renormalizable Lagrangian which preserves two independent parities such as
$\phi_A \rightarrow - \phi_A$ and $\phi_B \rightarrow - \phi_B$.

We first consider the second term in the parenthesis of \eqref{decomposition}.
In actual evaluations of \eqref{decomposition}, the contribution of this term can be extracted from the contribution of the first term
by taking $n = 1$, but to review the calculation of the second term is a good preparation for considering the contribution of the first term.
Decomposing the action of the total system into the noninteracting part and the interacting part, 
$\mathcal{S}^{(\beta)} = \mathcal{S}^{(\beta)}_0 + \mathcal{S}^{(\beta)}_I$,
we obtain
\begin{equation}
\begin{split}
\log \frac{Z_{\rm tot}^{(\beta)}}{Z_{A,0}^{(\beta)}  \, Z_{B,0}^{(\beta)}} = \log \left(  1 + \sum_{l=1}^\infty \frac{1}{l!} 
\frac{\int \mathcal{D} \phi_A \mathcal{D}  \phi_B \, e^{-\mathcal{S}_0^{(\beta)} } (\mathcal{S}_I^{(\beta)} )^l }{\int \mathcal{D}
\phi_A \mathcal{D}  \phi_B \, e^{-\mathcal{S}_0^{(\beta)} }} \right) . \label{Zbetaexpansion1}
\end{split}
\end{equation}
This has the same form with the interaction contributions in the usual finite-temperature perturbation theory.
In each term of the summation, a power of $\mathcal{S}^{(\beta)}_I$ is averaged over the unperturbed ensemble.

Following Ref.~\cite{Kapusta:2006pm}, we summarize Feynman rules in the diagrammatic calculations
of the contributions \eqref{Zbetaexpansion1}.
First, define the propagators of $\phi_A$, $\phi_B$ as
\begin{equation}
\begin{split}
D^{(\beta)}_\alpha (\tau , \bm{x}) &= \frac{1}{\beta} \sum_{m= -\infty}^{\infty} \int \frac{d^d p}{(2 \pi)^d} \,
e^{i (\omega_m \tau + \bm{p} \cdot \bm{x})} \widetilde{D}^{(\beta)}_\alpha (\omega_m , \bm{p}) \\[1ex]
&= \frac{1}{\beta} \sum_{m= -\infty}^{\infty} \int \frac{d^d p}{(2 \pi)^d} \,
\frac{e^{i (\omega_m \tau + \bm{p} \cdot \bm{x})}}{\omega_m^2 + \bm{p}^2 + M_\alpha^2} \, , \label{ABpropagator}
\end{split}
\end{equation}
where $\widetilde{D}_\alpha^{(\beta)} (\omega_m, \bm{p})$ $(\alpha = A, B)$ are the propagators in momentum space
and $\omega_m = 2 \pi m T$ $( m \in \bf{Z})$.
$M_{A,B}$ is the mass of $\phi_{A,B}$.
The position-space rules for associating equations with pieces of diagrams are 
\begin{figure}[H]
\begin{center}
   \includegraphics[clip, width=9.3cm]{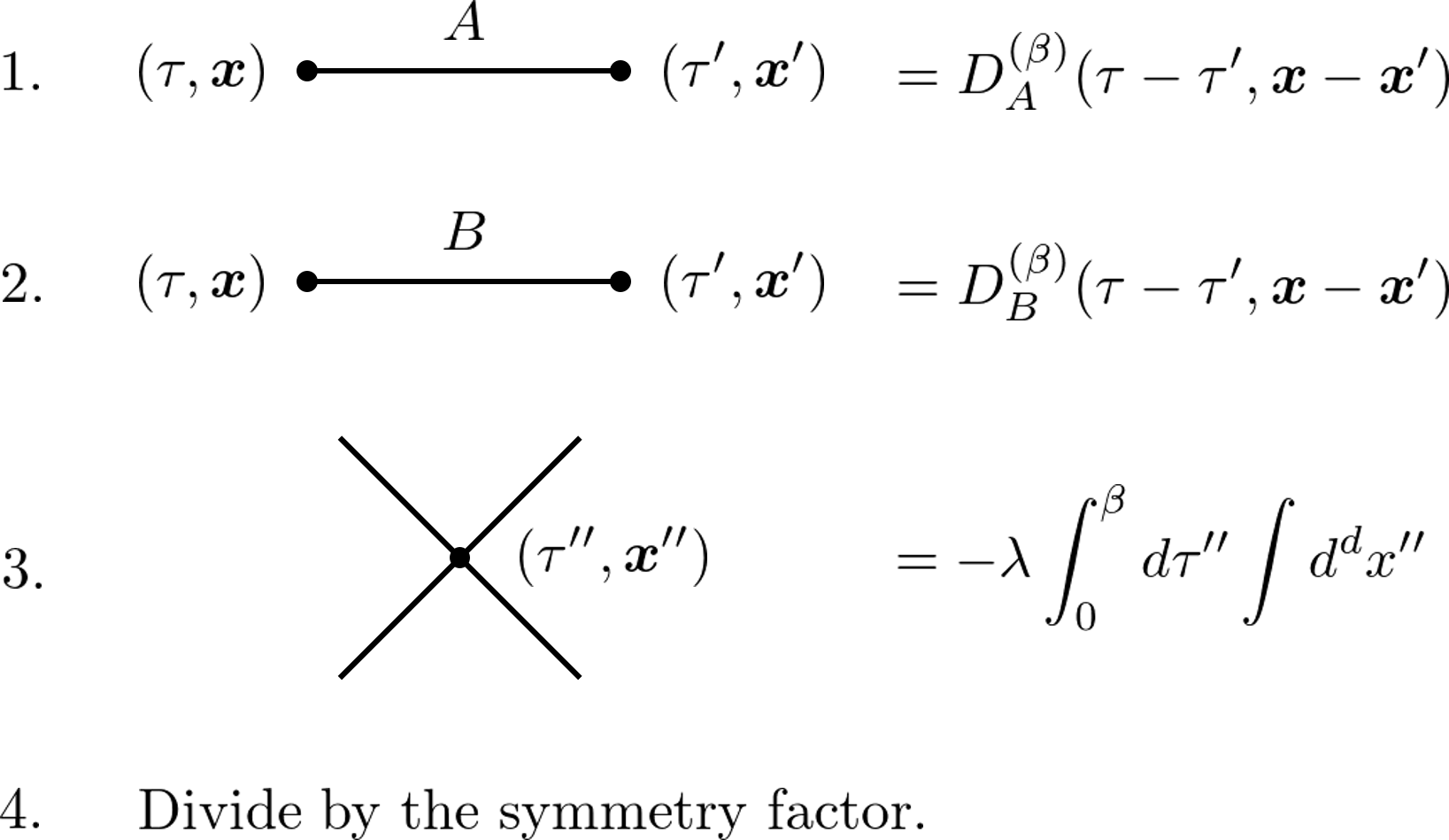}
\vspace{-0.5cm}
\end{center}
\end{figure}
\noindent The rules for the self-interaction terms in \eqref{scalarint} can be obtained by replacing $\lambda$ with $\lambda_A$ and
$\lambda_B$ in the rule~$3$.
We draw all topologically inequivalent diagrams to a given order of the perturbation theory.
If we label all the possible connected diagrams by $C$ and the sum of their contributions by $V_C$,
$N$ disconnected pieces of $C$ contribute as $(V_C)^N / N!$ due to the symmetry factor.
Then, we can easily see that only connected diagrams contribute to \eqref{Zbetaexpansion1}
by the exponentiation of disconnected diagrams.

In momentum space, four lines meet at a vertex and $(\tau'' , \bm{x}'')$-dependent factors of a diagram are given by
\begin{figure}[H]
\hspace{-0.3cm}
 \begin{minipage}{0.5\hsize}
  \begin{center}
   \includegraphics[clip, width=5.4cm]{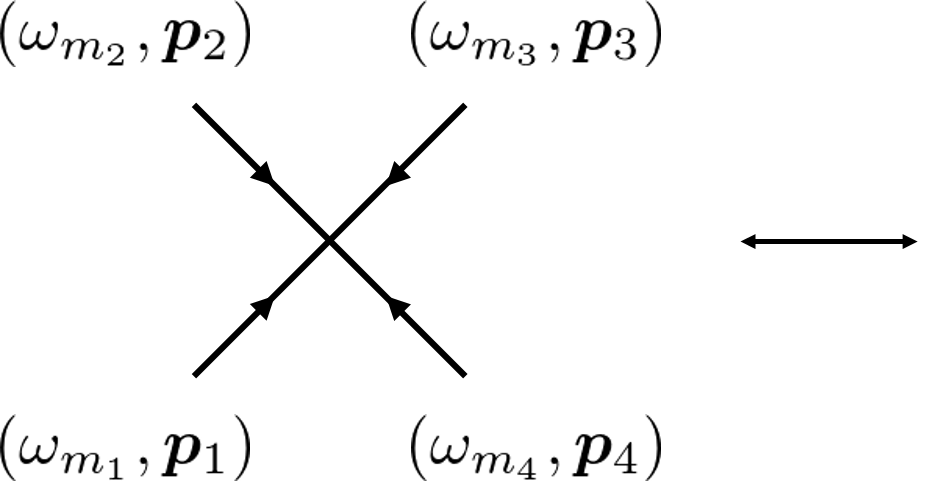}
  \end{center}
 \end{minipage}
\hspace{-0.7cm}
 \begin{minipage}{0.5\hsize}
  \begin{center}
   \includegraphics[clip, width=7.6cm]{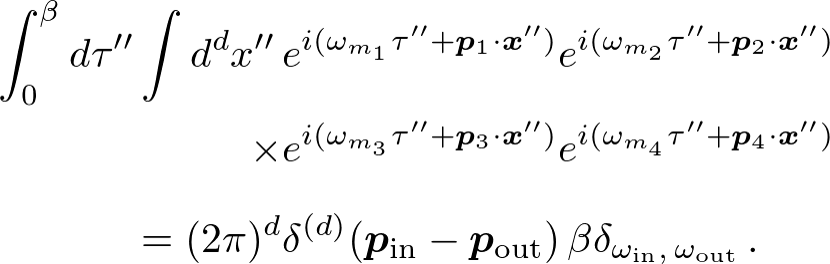}
  \end{center}
 \end{minipage}
\end{figure}
\noindent Here, we have defined the sum of ingoing momenta to the vertex and that of outgoing momenta as
$\bm{p}_{\rm in}$ and  $\bm{p}_{\rm out}$ respectively.
In the above case, $\bm{p}_{\rm in} = \bm{p}_1 + \bm{p}_2 + \bm{p}_3 + \bm{p}_4$ and
$\bm{p}_{\rm out} = 0$. $\omega_{\rm in}$ and $\omega_{\rm out}$ are defined in the same way.
We now summarize momentum-space Feynman rules as follows:
\begin{enumerate}
  \item For each propagator of $\phi_{A,B}$, assign a factor
$\frac{1}{\beta} \sum_m \int \frac{d^d p }{(2 \pi)^d} \, \widetilde{D}^{(\beta)}_{A,B} (\omega_m , \bm{p})$.
  \item Include a factor $- \lambda (2 \pi)^d \delta^{(d)} (\bm{p}_{\rm in} - \bm{p}_{\rm out}) \,
\beta \delta_{\omega_{\rm in} , \, \omega_{\rm out}}$ for each vertex.
  \item Divide by the symmetry factor.
\end{enumerate}
In the rule~$2$, $\lambda$ is replaced to $\lambda_{A,B}$ for a vertex of $\frac{\lambda_{A,B}}{4!} \phi_{A,B}^4$.
We draw all topologically inequivalent diagrams to a given order of the perturbation theory.
After the calculation of each diagram, an overall factor $\beta (2 \pi)^d \delta^{(d)} (0) = \beta V$ always appears.

We now consider the first term in the parenthesis of \eqref{decomposition}.
As in the case of the second term, let us first decompose the action of the total system
$\widetilde{\mathcal{S}}^{\, (n\beta)}$ (the tilde on $\mathcal{S}$ represents
the difference of periodicity between $\phi_A$ and $\phi_B$)
into the noninteracting part and the interacting part,
\begin{equation}
\begin{split}
\widetilde{\mathcal{S}}^{\, (n\beta)} & = \, \widetilde{\mathcal{S}}^{\, (n\beta)}_0 + \widetilde{\mathcal{S}}^{\, (n\beta)}_I \\[1ex]
&= \, \int_{0}^{n\beta} d\tau \int d^d x
\, \mathcal{L}_{A, 0} ( \phi_A ) \, + \, \sum_{j=1}^{n} \int_{(j-1)\beta}^{j\beta -\epsilon} d\tau \int d^d x
\, \mathcal{L}_{B, 0} (\phi_B^{(j)} )  \\
&\qquad + \, \sum_{j=1}^{n} \int_{(j-1)\beta}^{j\beta -\epsilon} d\tau \int d^d x
\, \mathcal{L}_I ( \phi_A , \phi_B^{(j)} ) \, .
\end{split}
\end{equation}
The terms in the first line of the right-hand side denote the free action of $\phi_A$ and $\phi_B^{(j)}$, each of which
is given by \eqref{freescalar}.
The last term is the interacting part.
It is important to note that we can consider this action as the theory of $(n+1)$ scalar fields, $\phi_A$ and $\phi_B^{(j)} \, (j = 1, \cdots, n)$
although periodicity of $\tau$ in the interacting part has an unusual structure.
We then expand the first term in the parenthesis of \eqref{decomposition} in a power series of $\widetilde{\mathcal{S}}^{\, (n\beta)}_I$,
\begin{equation}
\begin{split}
\log \frac{\widetilde{Z}_{\rm tot}^{(n\beta)}}{Z_{A,0}^{(n\beta)}  \, (Z_{B,0}^{(\beta)})^n} = \log \left(  1 + \sum_{l=1}^\infty \frac{1}{l!} 
\frac{\int \mathcal{D} \phi_A \mathcal{D}  \phi_B \, e^{-\widetilde{\mathcal{S}}_0^{\, (n\beta)} }
(\widetilde{\mathcal{S}}_I^{\, (n\beta)} )^l }{\int \mathcal{D}
\phi_A \mathcal{D}  \phi_B \, e^{-\widetilde{\mathcal{S}}_0^{\, (n\beta)} }} \right) . \label{Zbetaexpansion}
\end{split}
\end{equation}
In each term of the sum, a power of $\widetilde{\mathcal{S}}^{\, (n\beta)}_I$ is averaged over the unperturbed ensemble.

The perturbation theory can be formulated in a similar way as we did in the discussion of
the second term in the parenthesis of \eqref{decomposition}, but there are several important differences.
First, taking into account periodicity in imaginary time of $\phi_A$ and multiplicity of $\phi_B$,
the propagators of $\phi_A$, $\phi_B$ have to be changed into
\begin{equation}
\begin{split}
&D^{(n\beta)}_A (\tau , \bm{x}) = \frac{1}{n\beta} \sum_{m= -\infty}^{\infty} \int \frac{d^d p}{(2 \pi)^d} \,
e^{i (\widetilde{\omega}_m \tau + \bm{p} \cdot \bm{x})} \widetilde{D}^{(n\beta)}_A (\widetilde{\omega}_m , \bm{p}) 
\qquad (0 \leq \tau < n \beta) \, , \\[1ex]
&D^{(\beta)}_{B, \, j} (\tau , \bm{x}) = \frac{1}{\beta} \sum_{m= -\infty}^{\infty} \int \frac{d^d p}{(2 \pi)^d} \,
e^{i (\omega_m \tau + \bm{p} \cdot \bm{x})} \widetilde{D}^{(\beta)}_{B, \, j} (\omega_m , \bm{p}) 
\qquad ((j-1) \beta \leq \tau < j \beta) \, , \label{propagatornbeta}
\end{split}
\end{equation}
where $\widetilde{D}_A^{(n\beta)} (\widetilde{\omega}_m, \bm{p})$ and $\widetilde{D}^{(\beta)}_{B, \, j} (\omega_m , \bm{p}) $
are the propagators of $\phi_A$ and $\phi_B^{(j)}$ in momentum space
and $\widetilde{\omega}_m = 2 \pi m T/n$ $( m \in \bf{Z})$.
Note that the propagator of $\phi_A$ is defined in $0 \leq \tau < n \beta$ while the propagator of $\phi_B^{(j)}$
is defined in $(j-1) \beta \leq \tau < j \beta$.
Position-space Feynman rules are now presented as
\begin{figure}[H]
\begin{center}
   \includegraphics[clip, width=9.7cm]{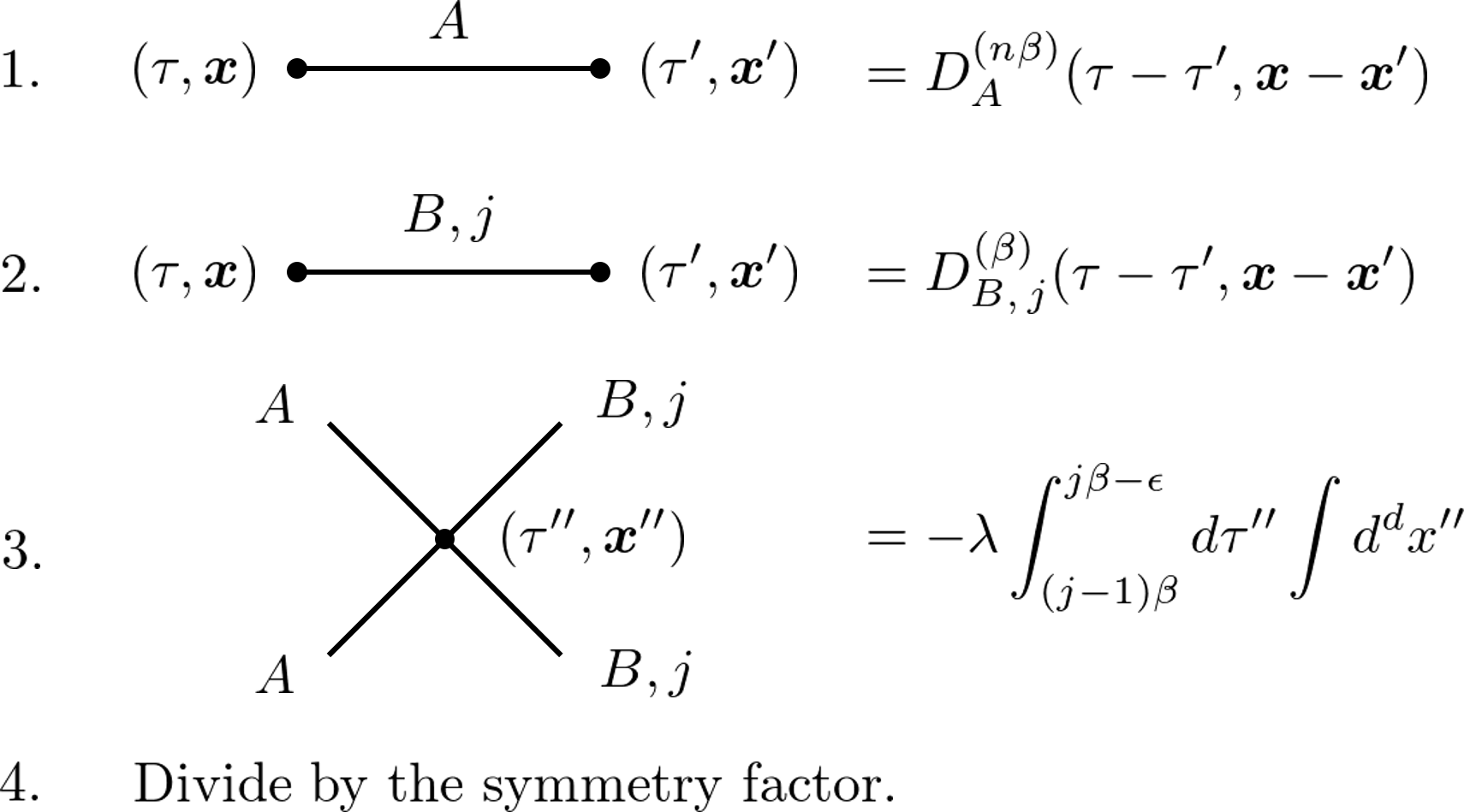}
\vspace{-0.5cm}
\end{center}
\end{figure}
\noindent The rule for the self-interaction term $\frac{\lambda_B}{4!} (\phi_B^{(j)} )^4$
can be obtained by replacing $\lambda$ with $\lambda_B$ in the rule~$3$. 
As for $\frac{\lambda_A}{4!} \phi_A^4$, 
the interval of $\tau''$ integration is changed into $0 \leq \tau'' < n \beta$
as well as replacing $\lambda$ with $\lambda_A$ in the rule~$3$. 
Note that 
there exists a line for each $\phi_B^{(j)}$ 
and lines with different $j$s do not directly connect with each other.

\begin{figure}[!t]
\vspace{-0.5cm}
 \begin{minipage}{0.5\hsize}
  \begin{center}
   \includegraphics[clip, width=4.5cm]{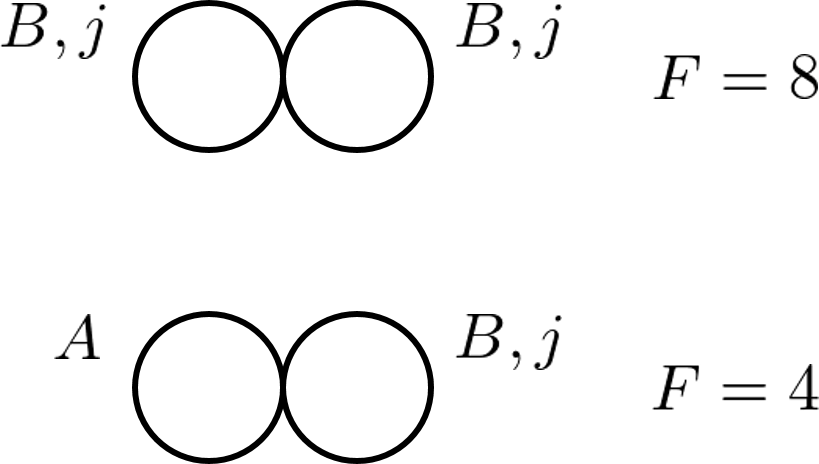}
  \end{center}
 \end{minipage}
\hspace{-0.7cm}
 \begin{minipage}{0.5\hsize}
\vspace{-0.4cm}
  \begin{center}
   \includegraphics[clip, width=5.4cm]{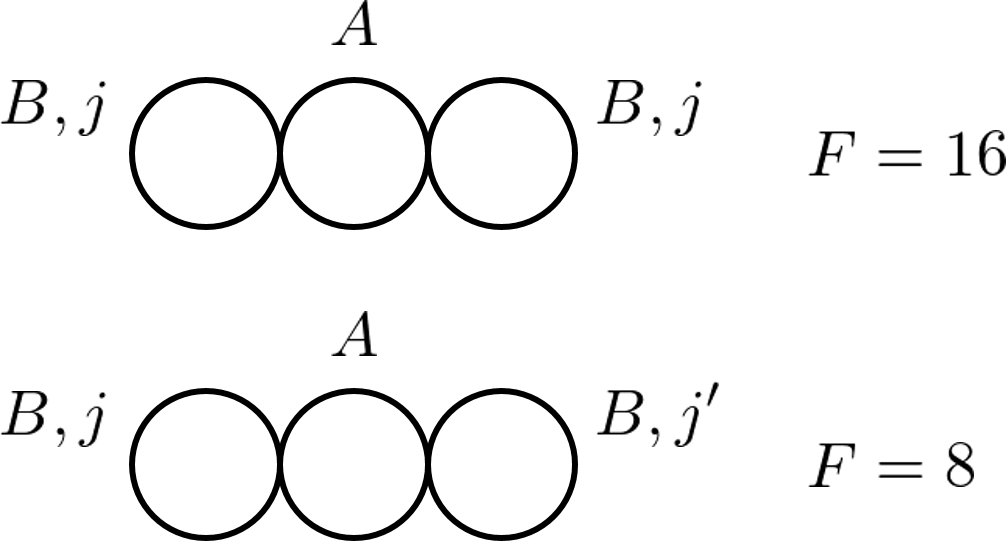}
  \end{center}
 \end{minipage}
\vspace{0.5cm}
  \caption{Examples of Feynman diagrams for the evaluation of the first term in the parenthesis of \eqref{decomposition}.
The left two diagrams give the leading order contributions while the right two diagrams contribute at the higher order.
In the bottom right diagram, we take $j \neq j'$.
The symmetry factor $F$ of each diagram is also shown.}
  \label{fig:diagram}
\end{figure}

We draw all topologically inequivalent diagrams to a given order of the perturbation theory.
Some examples of the diagrams are shown in Figure~\ref{fig:diagram}
where the symmetry factor $F$ of each diagram is also presented.
In the top left diagram, the number of pairing lines is $3$. Dividing this number by $4!$ in the interaction term,
we obtain $F=8$. 
For the other diagrams, $F$ can be counted in the same way.
It can also be understood from the symmetry of a diagram.
For the right two diagrams, $F$s are different by a factor of $2$ because the bottom diagram is asymmetric under the interchange of
loops at the two ends.
This difference originally comes from $(\widetilde{\mathcal{S}}_I^{\, (n\beta)} )^2$ which contains
two $(\phi_A \phi_B^{(j)})^2 (\phi_A \phi_B^{(j')})^2$ for $j \neq j'$.
As in the case of the calculation of the second term in the parenthesis of \eqref{decomposition},
if we label all the possible connected diagrams by $\widetilde{C}$ and the sum of their contributions by $V_{\widetilde{C}}$,
$N$ disconnected pieces of $\widetilde{C}$ contribute as $(V_{\widetilde{C}})^N / N!$ due to the symmetry factor.
Then, only connected diagrams contribute to \eqref{Zbetaexpansion}
by the exponentiation of disconnected diagrams.

There is 
another
important difference from the calculation of the second term in the parenthesis of \eqref{decomposition}
when we move to momentum space.
In the present case, $(\tau'' , \bm{x}'')$-dependent factors at a vertex of $\frac{\lambda}{4} \phi_A^2 \phi_B^2$ are given by
\begin{figure}[H]
\hspace{-0.6cm}
 \begin{minipage}{0.5\hsize}
  \begin{center}
   \includegraphics[clip, width=5.5cm]{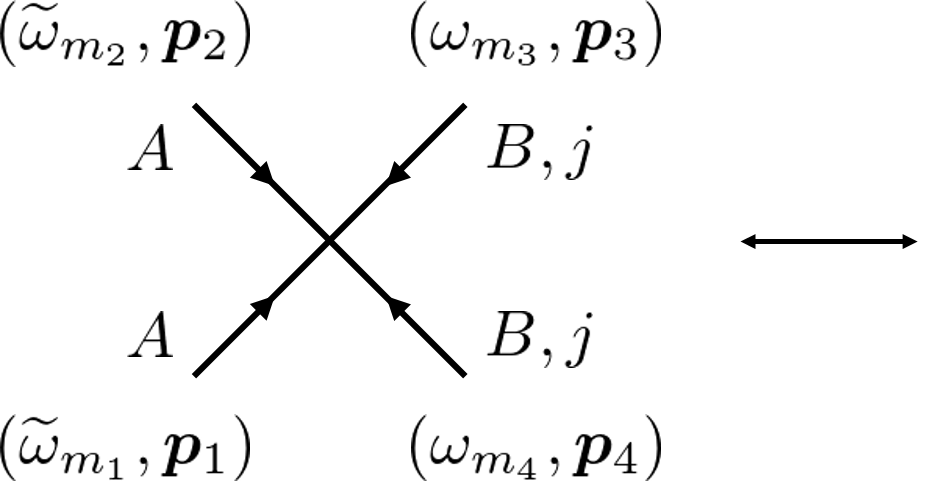}
  \end{center}
 \end{minipage}
\hspace{-1cm}
 \begin{minipage}{0.5\hsize}
  \begin{center}
   \includegraphics[clip, width=8.5cm]{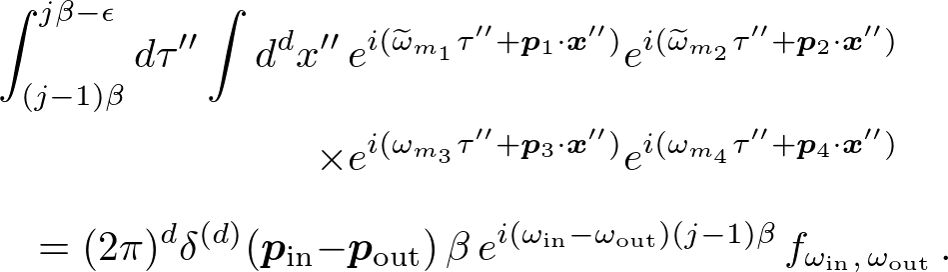}
  \end{center}
 \end{minipage}
\end{figure}
\noindent Here, $\omega_{\rm in} = \widetilde{\omega}_{m_1} + \widetilde{\omega}_{m_2} + \omega_{m_3} + \omega_{m_4}$
is the sum of ingoing energies to the vertex
and $\omega_{\rm out} = 0$ is that of outgoing energies.
The function $f$ is defined as
\begin{equation}
\begin{split}
f_{\omega_{\rm in}, \, \omega_{\rm out}} \equiv \frac{1}{\beta}\int_{0}^{\beta}
d\tau'' \, e^{i(\omega_{\rm in}-\omega_{\rm out})\tau''} 
=
\begin{cases}
\, 1 & \text{for } \omega_{\rm in} = \omega_{\rm out} \\[1ex]
\,  \frac{e^{i  (\omega_{\rm in}-\omega_{\rm out}) \beta} -1  }
{i (\omega_{\rm in}-\omega_{\rm out}) \beta} 
& \text{for } \omega_{\rm in} \neq \omega_{\rm out} \, . \label{fl}
\end{cases}
\end{split}
\end{equation}
Note that the factor $e^{i  (\omega_{\rm in}-\omega_{\rm out}) \beta}$ is not equal to $1$ in general when $\omega_{\rm in} \neq \omega_{\rm out} $
because the energy of $\phi_A$ has the form $\widetilde{\omega}_m = 2 \pi m T /n$ with $n \neq 1$.
This means that in the calculation of the first term in the parenthesis of \eqref{decomposition},
energy is not necessarily conserved at a vertex of $\frac{\lambda}{4} \phi_A^2 \phi_B^2$ 
because of the difference of periodicity. 
On the other hand, for a vertex of $\frac{\lambda_A}{4!} \phi_A^4$ or $\frac{\lambda_B}{4!} \phi_B^4$, energy conservation is respected as usual.

We now summarize momentum-space Feynman rules as follows:
\begin{enumerate}
  \item For each propagator of $\phi_{A}$, assign a factor
$\frac{1}{n\beta} \sum_m \int \frac{d^d p }{(2 \pi)^d} \, \widetilde{D}^{(n \beta)}_{A} (\widetilde{\omega}_m , \bm{p})$.
  \item For each propagator of $\phi_{B}^{(j)}$, assign a factor
$\frac{1}{\beta} \sum_m \int \frac{d^d p }{(2 \pi)^d} \, \widetilde{D}^{(\beta)}_{B, \, j} (\omega_m , \bm{p})$.
  \item Include a factor $- \lambda (2 \pi)^d \delta^{(d)} (\bm{p}_{\rm in} - \bm{p}_{\rm out}) \,
\beta \, e^{i (\omega_{\rm in} - \omega_{\rm out}) (j-1) \beta }f_{\omega_{\rm in} , \, \omega_{\rm out}}$
for each vertex of $\frac{\lambda}{4} (\phi_A \phi_B^{(j)})^2$.
  \item Include a factor $- \lambda_{A} (2 \pi)^d \delta^{(d)} (\bm{p}_{\rm in} - \bm{p}_{\rm out}) \,
n \beta \delta_{\omega_{\rm in} , \, \omega_{\rm out}}$ for each vertex of $\frac{\lambda_{A}}{4!} \phi_{A}^4$.
  \item Include a factor $- \lambda_{B} (2 \pi)^d \delta^{(d)} (\bm{p}_{\rm in} - \bm{p}_{\rm out}) \,
\beta \delta_{\omega_{\rm in} , \, \omega_{\rm out}}$ for each vertex of 
$\frac{\lambda_{B}}{4!} \left( \phi_{B}^{(j)} \right)^4$.
  \item Divide by the symmetry factor.
\end{enumerate}
Note that there is an extra factor $n$ in the rule~$4$ compared to the rule~$5$.
We draw all topologically inequivalent diagrams to a given order of the perturbation theory. 
In particular, 
we should take a summation of the index $j$. 
In the next section, we will 
present an explicit
calculation of the leading order correction to the thermodynamic entropy in an ideal Bose gas
by using these Feynman rules.
In section~\ref{QED} and \ref{Yukawa}, we will consider QED and the Yukawa theory, respectively.
Rules for these cases will be summarized there.

\section{Coupled $\phi^4$ theory}\label{phi4}

We here perform an explicit calculation of Feynman diagrams in the scalar-scalar system presented in the previous section
and find the leading order correction to the thermodynamic entropy in an ideal Bose gas.
Nonzero contributions to the second term in \eqref{decomposition} start from two-loop diagrams.
Although we encounter divergent loop integrals in the calculation of these diagrams,
it is shown that the divergence is renormalized correctly 
by adding counterterms that are equal to those in the usual zero-temperature field theory.

We start with presenting the total Lagrangian of the coupled $\phi^4$ theory again,
\begin{equation}
\begin{split}
\mathcal{L} (\phi_A, \phi_B) &= \mathcal{L}_{0} + \mathcal{L}_{I} + \mathcal{L}_{\rm counter} \, ,\\[1.5ex]
\mathcal{L}_{0} (\phi_A, \phi_B) &=\frac{1}{2} \left[ (\partial_{\mu}\phi_{A})^2 +M_{A}^{2} \phi_{A}^{2}  \right]
+\frac{1}{2} \left[ (\partial_{\mu}\phi_{B})^2 +M_{B}^{2} \phi_{B}^{2}  \right] , \\[1.5ex]
\mathcal{L}_{I} (\phi_A, \phi_B) &=\frac{\lambda_A}{4!} \phi_{A}^{4} +\frac{\lambda_B}{4!} \phi_{B}^{4}+\frac{\lambda}{4}\phi_{A}^{2} \phi_{B}^{2} \, , \\[1.5ex]
\mathcal{L}_{\rm counter} (\phi_A, \phi_B)&=\frac{1}{2} \left[ \delta_{Z_A} (\partial_{\mu}\phi_{A})^2
+\delta_{M_A} \phi_{A}^{2} \right] +\frac{1}{2} \left[ \delta_{Z_B} (\partial_{\mu}\phi_{B})^2
+\delta_{M_B} \phi_{B}^{2} \right] \\[1ex]
&\quad+\frac{\delta_{\lambda_A}}{4!} \phi_{A}^{4} +\frac{\delta_{\lambda_B}}{4!} \phi_{B}^{4}+\frac{\delta_\lambda}{4}\phi_{A}^{2} \phi_{B}^{2} \, , \\[1.5ex]
\end{split}
\end{equation}
where we have defined $(\partial_{\mu}\phi)^2 \equiv (\partial_\tau \phi)^2 + (\nabla \phi)^2$.
$\mathcal{L}_{0}$ is the noninteracting part of the Lagrangian and $\mathcal{L}_{I}$ is the interacting part.
$\mathcal{L}_{\rm counter}$ denotes the counterterms that cancel divergence.
Since the counterterms corresponding to the interaction terms $\delta_{\lambda_A}$, 
$\lambda_{\delta_B}$, and $\delta_\lambda$ are
relevant only for the next-to-leading and higher order corrections,
we can neglect these terms in the following calculation.

\begin{figure}[!t]
\vspace{-0.5cm}
\hspace{-0.5cm}
 \begin{subfigure}{0.18\columnwidth}
\vspace{0.1cm}
\centering
   \includegraphics[clip, width=2.6cm]{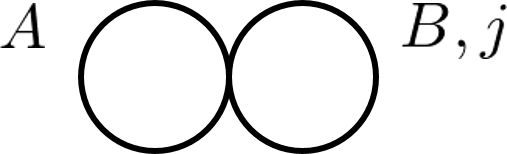}
\vspace{0.6cm}
\caption{}
\label{scalar(a)}
\end{subfigure}
\hspace{0.3cm}
\centering
 \begin{subfigure}{0.18\columnwidth}
\vspace{0.1cm}
\centering
   \includegraphics[clip, width=2.3cm]{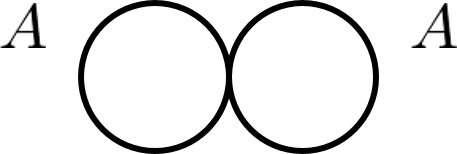}
\vspace{0.6cm}
\caption{}
\label{scalar(b)}
\end{subfigure}
\hspace{0.2cm}
 \begin{subfigure}{0.18\columnwidth}
\vspace{0.1cm}
\centering
   \includegraphics[clip, width=2.9cm]{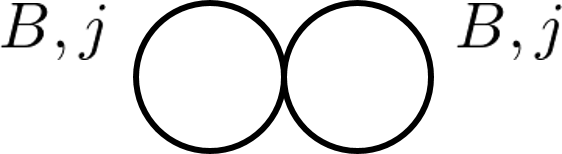}
\vspace{0.1cm}
\caption{}
\label{scalar(c)}
\end{subfigure}
\hspace{0.2cm}
 \begin{subfigure}{0.18\columnwidth}
\vspace{0.1cm}
\centering
   \includegraphics[clip, width=1.3cm]{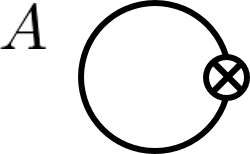}
\vspace{0.7cm}
\caption{}
\label{scalar(d)}
\end{subfigure}
\vspace{-0.2cm}
\hspace{-0.4cm}
\centering
 \begin{subfigure}{0.18\columnwidth}
\vspace{0.1cm}
\centering
   \includegraphics[clip, width=1.6cm]{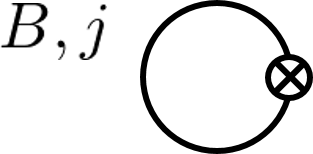}
\vspace{0.7cm}
\caption{}
\label{scalar(e)}
\end{subfigure}
\vspace{0.3cm}
  \caption{Feynman diagrams which contribute to the first term in the parenthesis of \eqref{decomposition} at the leading order.
The last two diagrams come from the counterterms.}
 \label{fig:phi4diagram}
\end{figure}

We now calculate the entanglement (R\'{e}nyi) entropy of the coupled $\phi^4$ theory in $3+1$ dimensions by using the formula \eqref{decomposition}
and Feynman rules discussed in the previous section.
Figure~\ref{fig:phi4diagram} shows Feynman diagrams which contribute to the first term in the parenthesis of \eqref{decomposition} at the leading order.
The last two diagrams come from the counterterms. The rules for the counterterms are easily obtained from the above Lagrangian.
The diagram~\ref{scalar(a)} contains a $\phi_A$ loop and a $\phi_B^{(j)} \, (j = 1, \cdots , n)$ loop, 
where 
the symmetry factor is $F =4$.
Using momentum-space Feynman rules, we obtain
\begin{equation}
\begin{split}
(\text{Fig.~\ref{scalar(a)}} ) = &\,\frac{1}{4} \, \sum_{j=1}^{n} \, \biggl( \frac{1}{n\beta} \sum_{m_A} \int \frac{d^3 p_A}{(2\pi)^3} \,
\widetilde{D}_{A}^{(n\beta)}(\widetilde{\omega}_{{m_A}}, \bm{p}_A) \biggr) \\
&\times \biggl( \frac{1}{\beta} \sum_{m_{B}} \int \frac{d^3 p_{B}}{(2\pi)^3} \, \widetilde{D}_{B, \, j}^{(\beta)}(\omega_{m_{B}}, \bm{p}_{B}) \biggr) \times
(- \lambda)  (2\pi)^{3} \delta^3(0) \, \beta \\[1ex]
=  &-\frac{\lambda}{4} \, n\beta V \, D_{A}^{(n\beta)} (0, 0) \, D_{B}^{(\beta)} (0, 0)  \, , \label{resulta}
\end{split}
\end{equation}
where we used ${\bm p}_{\rm in} = \bm{p}_A + \bm{p}_B$, 
$\bm{p}_{\rm out} = \bm{p}_A + \bm{p}_B$, 
$\omega_{\rm in} = \widetilde{\omega}_{m_A}  + \omega_{m_B}$, 
$\omega_{\rm out} = \widetilde{\omega}_{m_A}  + \omega_{m_B}$, 
and $(2 \pi)^3 \delta^3 (0) = V$. 
Here, $\widetilde{\omega}_{m_A} = 2 \pi m_A T / n$, $\omega_{m_B} = 2 \pi m_B T$. 
In this diagram, the factor $e^{i  (\omega_{\rm in}-\omega_{\rm out}) (j-1) \beta} \, f$, 
which comes from the rule~$3$ of the Feynman rules, 
is just equal to $1$ and energy is (accidentally) conserved. 
The diagram~\ref{scalar(b)} contains two $\phi_A$ loops, where the symmetry factor is now given by $F =8$.
In the same way as the diagram~\ref{scalar(a)}, we can evaluate this diagram as
\begin{equation}
\begin{split}
(\text{Fig.~\ref{scalar(b)}} ) &= -\frac{\lambda_A}{8} \, n\beta V \left( D_{A}^{(n\beta)} (0,0) \right)^{2} . \label{resultb}
\end{split}
\end{equation}
The diagram~\ref{scalar(c)} contains two $\phi_B^{(j)} $ loops (note that both of loops have the same $j$).
The symmetry factor is also $F =8$. 
We find
\begin{equation}
\begin{split}
(\text{Fig.~\ref{scalar(c)}} ) &= -\frac{\lambda_B}{8} \, n\beta V \left( D_{B}^{(\beta)} (0,0) \right)^{2}  . \label{resultc}
\end{split}
\end{equation}
Next, we evaluate the diagrams which contain the counterterms.
The diagram~\ref{scalar(d)} with a $\phi_A$ loop leads to 
\begin{equation}
\begin{split}
(\text{Fig.~\ref{scalar(d)}} ) = -\frac{1}{2} \, n\beta V \, \biggl[ &\, \delta_{M_A} \, D_{A}^{(n\beta)} (0,0) \\[1ex]
&+\frac{1}{n\beta} \, \sum_{m_A} \int \frac{d^3 p_A}{(2\pi)^3} \, \widetilde{D}_{A}^{(n\beta)}(\widetilde{\omega}_{m_A}, \bm{p}_A)
\, \delta_{Z_A} \left(\widetilde{\omega}_{m_A}^2+p_A^2 \right) \biggr]  \, , \label{resultd}
\end{split}
\end{equation}
where we include the symmetry factor $F=2$.
In the same way, the diagram~\ref{scalar(e)} with a $\phi_B^{(j)}$ loop is calculated as
\begin{equation}
\begin{split}
(\text{Fig.~\ref{scalar(e)}} ) = -\frac{1}{2} \, n\beta V \, \biggl[ &\, \delta_{M_B} \, D_{B}^{(\beta)} (0,0) \\[1ex]
&+\frac{1}{\beta} \, \sum_{m_B} \int \frac{d^3 p_B}{(2\pi)^3} \, \widetilde{D}_{B}^{(\beta)}
(\omega_{m_B}, \bm{p}_B) \, \delta_{Z_B} \left( \omega_{m_B}^2+p_B^2 \right) \biggr] \, . \label{resulte}
\end{split}
\end{equation}
The first term in the parenthesis of \eqref{decomposition} is given by the sum of the contributions from Figure~\ref{scalar(a)}-\ref{scalar(e)}.

The contributions from the diagrams~\ref{scalar(a)}-\ref{scalar(c)} contain the propagators such as
$D_{A}^{(n\beta)} (0,0)$, $D_{B}^{(\beta)} (0,0)$ which are given by divergent integrals.
We need to set renormalization conditions and determine the parameters in the counterterms $\mathcal{L}_{\rm counter}$ 
so that 
the divergence is canceled by the contributions from the diagrams with the counterterms~\ref{scalar(d)}, \ref{scalar(e)}. 
The divergence arises in the UV region and comes from the 
zero-temperature part of the propagators. 
To see this, 
let us first decompose the propagators $D_{A}^{(n\beta)} (0,0)$, $D_{B}^{(\beta)} (0,0)$ into the $T=0$ part and $T \neq 0$ part 
by using the following formula for the frequency sums
\cite{Kapusta:2006pm}:
\begin{equation}
\begin{split}
\frac{1}{\beta} \sum_{m = - \infty}^{\infty} \mathcal{F} (p_0 = i \omega_m = 2 \pi m T i) = & \, \frac{1}{2 \pi i} \int_{- i \infty}^{i \infty} dp_0 \,
\frac{1}{2} \left[ \mathcal{F} (p_0) + \mathcal{F} (-p_0)  \right] \\[1ex]
&+\frac{1}{2 \pi i} \int_{- i \infty + \epsilon}^{i \infty + \epsilon} dp_0 \,
 \left[ \mathcal{F} (p_0) + \mathcal{F} (-p_0)  \right] \frac{1}{e^{\beta p_0} -1} \, , \label{Fformula}
\end{split}
\end{equation}
where $\mathcal{F} (p_0)$ is some function which has no singularities along the imaginary $p_0$ axis.
For the propagator $D_{B}^{(\beta)} (0,0)$, we take $\mathcal{F} (p_0) = \int \frac{d^3 p}{(2\pi)^3} (- p_0^2 + \bm{p}^2 + M_{B}^2)^{-1}$.
In this case, we can perform a contour integral with a residue at $p_0 = \omega \equiv \sqrt{\bm{p}^2 + M_{B}^2}$
in the second term of \eqref{Fformula}.
In addition, we change the variable as $p_0 \rightarrow - i p_4$ in the first term.
Then, we obtain
\begin{equation}
\begin{split}
D_{B}^{(\beta)}(0,0) &=D_{B}^{\rm vac}+D_{B}^{\rm mat}(\beta)  \\[1ex]
&\equiv \int \frac{d^4 p }{(2\pi)^4} \, \frac{1}{p_4^2+ \bm{p}^2+M_B^2} + \int \frac{d^3 p }{(2\pi)^3} \,
\frac{1}{\omega}\frac{1}{e^{\beta \omega} -1} \, ,
\end{split}
\end{equation}
where $d^4 p = dp_4 d^3 p$.
In the same way, the propagator of $\phi_A$ can be decomposed into the $T=0$ part and $T \neq 0$ part: 
\begin{equation}
\begin{split}
D_{A}^{(n\beta)}(0,0) &=D_{A}^{\rm vac}+D_{A}^{\rm mat}(n\beta) \\[1ex]
&\equiv \int \frac{d^4 p }{(2\pi)^4} \, \frac{1}{p_4^2+\bm{p}^2+M_A^2} + \int \frac{d^3 p }{(2\pi)^3} \,
\frac{1}{\widetilde{\omega}}\frac{1}{e^{n\beta \widetilde{\omega}} -1} \, , 
\end{split}
\end{equation}
where $\widetilde{\omega} \equiv\sqrt{\bm{p}^2+M_A^2}$.
We can see that the $T=0$ part of the propagators $D_{A,B}^{\rm vac}$ is divergent
and has to be removed by renormalization 
while the $T \neq 0$ part is finite due to the exponential factor in the denominator.

To set renormalization conditions and determine the parameters in the counterterms, we consider the usual (Euclidean)
zero-temperature field theory and
compute the sum of all one-particle-irreducible (1PI) insertions
into the propagator as in the case of the usual perturbation theory in the $\phi^4$ theory. For the propagator of $\phi_B$, we find
\begin{figure}[H]
\vspace{0cm}
  \begin{center}
   \includegraphics[clip, width=11.2cm]{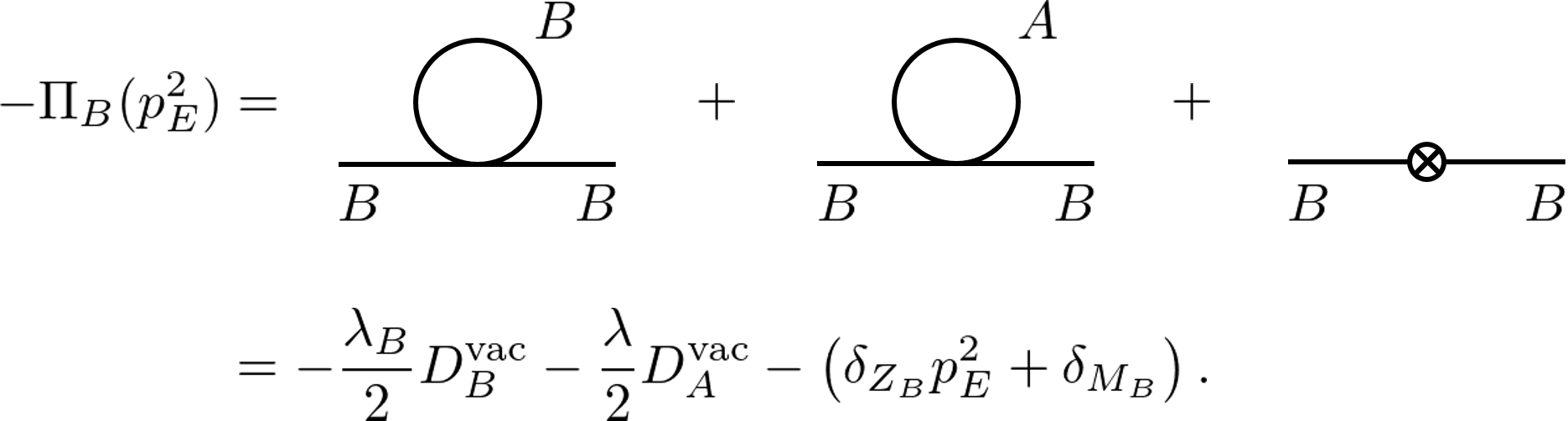}
  \end{center}
\vspace{-0.4cm}
\end{figure}
\noindent Here, we have defined $p_E^2=p_4^2+\bm{p}^2$.
We set renormalization conditions such that the pole in the full propagator given by geometric series of the sum of all 1PI insertions occur at
$p_E^2 = - M_B^2$ and have residue 1,
\begin{equation}
\begin{split}
\Pi_{B}(p_E^2=-M_B^2) =0 \, , \qquad  \frac{d}{dp_E^2} \Pi_{B} \biggr|_{p_E^2=-M_B^2} =0 \, .
\end{split}
\end{equation}
Inserting the expression of $\Pi_B$ computed above into these conditions, we can determine
the parameters in the counterterms as
\begin{equation}
\begin{split}
\delta_{Z_B}=0 \, , \qquad \delta_{M_B}=-\frac{\lambda_{B}}{2}  D_{B}^{\rm vac} -\frac{\lambda}{2}  D_{A}^{\rm vac} \, . \label{counterparaB}
\end{split}
\end{equation}
As is well known, $\delta_{Z_B}$ is trivial for the $\phi^4$ theory and wave function renormalization is not needed at the leading order.
In the same way, we also find the parameters in the counterterms corresponding to $\phi_A$,
\begin{equation}
\begin{split}
\delta_{Z_A}=0  \, , \qquad \delta_{M_A}=-\frac{\lambda_{A}}{2}  D_{A}^{\rm vac} -\frac{\lambda}{2}  D_{B}^{\rm vac} \, . \label{counterparaA}
\end{split}
\end{equation}
These choices of the parameters enable us to get a finite result of the entanglement (R\'{e}nyi) entropy.

Summing up the contributions from the diagrams~\ref{scalar(a)}-\ref{scalar(e)}, in which the propagators such as
$D_{A}^{(n\beta)} (0,0)$, $D_{B}^{(\beta)} (0,0)$
are decomposed into the $T=0$ part and $T \neq 0$ part, and using the relations of \eqref{counterparaB}
and \eqref{counterparaA},
the leading order contribution to the first term in the parenthesis of \eqref{decomposition} is obtained as
\begin{equation}
\begin{split}
 \log \frac{\tilde{Z}_{\rm tot}^{(n\beta)}}{Z_{A,0}^{(n\beta)}
 ( Z_{B,0}^{(\beta)} )^n } =n\beta V \biggl[ &-\frac{\lambda}{4} \left(D_{A}^{\rm vac}+D_{A}^{\rm mat}(n\beta) \right)
\left(D_{B}^{\rm vac}+D_{B}^{\rm mat}(\beta) \right) \\
& -\frac{\lambda_A}{8} \left(D_{A}^{\rm vac}+D_{A}^{\rm mat}(n\beta) \right)^2
-\frac{\lambda_B}{8} \left(D_{B}^{\rm vac}+D_{B}^{\rm mat}(\beta) \right)^2 \\[1ex]
&+\frac{1}{4} \left(\lambda_A D_{A}^{\rm vac}+\lambda D_{B}^{\rm vac} \right) \left(D_{A}^{\rm vac}+D_{A}^{\rm mat}(n\beta) \right) \\[1ex]
&+\frac{1}{4} \left(\lambda_B D_{B}^{\rm vac}+\lambda D_{A}^{\rm vac} \right) \left(D_{B}^{\rm vac}+D_{B}^{\rm mat}(\beta) \right) \biggr] \, .
\end{split}
\end{equation}
The contribution to the second term in the parenthesis of \eqref{decomposition} is obtained by taking $n=1$ in the above expression.
The  R\'{e}nyi entropy \eqref{decomposition} is then given by
\begin{equation}
\begin{split}
S_A^{(n)}-S_{A,0}^{(n)}
=\frac{n\beta V}{n-1} \, \biggl[\, &\frac{\lambda}{4}  \left( D_{A}^{\rm mat}(n\beta)-D_{A}^{\rm mat}(\beta) \right) D_{B}^{\rm mat}(\beta) \\[1ex]
&+\frac{\lambda_A}{8}  \left( D_{A}^{\rm mat}(n\beta)^2-D_{A}^{\rm mat}(\beta)^2 \right) \biggr] \, .
\end{split}
\end{equation}
Note that the divergent $T=0$ part of the propagators have vanished in the expression and the result is finite.
Taking the limit of $n \rightarrow 1$, we finally obtain the entanglement entropy \eqref{entanglemententropy} 
in the coupled $\phi^4$ theory as
\begin{equation}
\begin{split}
S_{A}-S_{A,0}
=-&\beta^2 V  \left[ \, \frac{\lambda}{4}  D_{B}^{\rm mat}(\beta) +\frac{\lambda_A}{4}  D_{A}^{\rm mat}(\beta) \right] \\[1ex]
&\times \int \frac{d^3 p }{(2\pi)^3} \left[ \frac{1}{e^{\beta \widetilde{\omega}} -1}
+ \left( \frac{1}{e^{\beta \widetilde{\omega}} -1} \right)^2 \, \right] . \label{phi4result}
\end{split}
\end{equation}
The term proportional to $\lambda_A$ is the quantum correction to the thermodynamic entropy existing even in the absence of the subsystem $B$.
The term proportional to $\lambda$ is the correction from the interaction between the subsystems.
In the high temperature limit, this expression of the entanglement entropy is approximately given by 
\begin{equation}
\begin{split}
S_{A} = V T^3 \left[  \frac{2 \pi^2}{45} - \frac{1}{12} \left( \frac{\lambda_A}{4!} \right)  - \frac{1}{12}
\left( \frac{\lambda}{4!} \right)  + \cdots \right] .  
\end{split}
\end{equation}
The first term is the usual entropy of an ideal Bose gas.
Note that the correction terms are by no means small when the coupling $\lambda$ or $\lambda_A$ is sufficiently strong, 
though there is a unitarity bound on the couplings such as $\lambda, \lambda_A \lesssim (4\pi)^2$.

Let us comment on the mutual information, $I(A, B) \equiv S_A + S_B  - S_{A+B} \geq 0$
(the inequality is always satisfied by the subadditivity of the entanglement entropy).
When we calculate the leading order correction to the thermodynamic entropy of the total system $S_{A+B}$,
we can obtain $I(A, B) = 0$ which means that there is no quantum entanglement at the present order in the coupled $\phi^4$ theory.
This is a special property of this theory at this order, which derives from the fact that the factor $f$ is trivial
in the contributions from the diagrams in Figure~\ref{fig:phi4diagram},
and not expected to be satisfied at higher orders.
There is still the correction coming from the interaction, 
which should be included even if the quantum entanglement is absent.

\section{Quantum electrodynamics}\label{QED}

In this section, we consider QED with one Dirac fermion and calculate the entanglement entropy of the fermion subsystem, tracing out the photon field
from the density matrix of the total system. 
Generalizations to more than one Dirac fermions and non-Abelian gauge theories such as QCD are straightforward.

Let us first summarize the total Lagrangian of QED in $3+1$ dimensions,
\begin{equation}
\begin{split}
\mathcal{L} (\psi, A_\mu) &= \mathcal{L}_0 + \mathcal{L}_I + \mathcal{L}_{\rm counter} \, , \\[1ex]
\mathcal{L}_0 (\psi, A_\mu)  &= \bar{\psi} \left( i \slashed{\partial} - M \right) \psi - \frac{1}{4} F^{\mu\nu} F_{\mu\nu}
- \frac{1}{2 \rho} \left( \partial^\mu A_\mu \right)^2 + \left( \partial^\mu \bar{C} \right) \left( \partial_\mu C \right) \, , \\[1ex]
\mathcal{L}_I (\psi, A_\mu)  &= - e \bar{\psi} \gamma^\mu \psi A_\mu   \, , \\[1ex]
\mathcal{L}_{\rm counter} (\psi, A_\mu)  &= \bar{\psi} \left( i \delta_\psi \slashed{\partial} - \delta_M \right) \psi
- \frac{1}{4} \delta_{\gamma} F^{\mu\nu} F_{\mu\nu}
- e \delta_e \bar{\psi} \gamma^\mu \psi A_\mu \, . \label{QEDLagrangian}
\end{split}
\end{equation}
Here, $\mathcal{L}_0$ is the noninteracting part of the Lagrangian of a Dirac fermion $\psi (t , \bm{x})$ and a photon field $A_\mu (t , \bm{x})$.
We have defined $\bar{\psi} \equiv \psi^\dagger \gamma^0$ and $\slashed{\partial} \equiv \gamma^\mu \partial_\mu$ as usual.
The second term of $\mathcal{L}_0$ denotes the photon kinetic term and $F_{\mu\nu} = \partial_\mu A_\nu - \partial_\nu A_\mu$
is the field strength.
The third term is the gauge fixing term.
Hereafter we choose the Feynman gauge and take $\rho = 1$.
In the fourth term, the ghost field $C$ is introduced to cancel contributions from 2 of the 4 degrees of freedom of the gauge field
to the free photon partition function, 
though it does not contribute to anything in QED. 
$\mathcal{L}_I$ is the usual interaction of QED and $e$ is the gauge coupling.
$\mathcal{L}_{\rm counter}$ denotes the counterterms that cancel divergence.
As in the case of the coupled $\phi^4$ theory, the term with $\delta_e$ is relevant only for the next-to-leading and higher order corrections
and we do not consider this term below.
In the following calculations, we assume that the chemical potential of the fermion field is zero for simplicity.
In cosmology, this is usually a good approximation 
because asymmetry is difficult to be generated as we can see 
from the fact that baryon asymmetry of the Universe is tiny, $\mu_B / T \sim 10^{-10} \ll 1$.
A generalization to the nonzero case is straightforward.

We now present Feynman rules to calculate the correction terms of \eqref{decomposition} in QED.
The rules for the second term in the parenthesis are the same as those of the ordinary finite-temperature field theory.
They are summarized in Ref.~\cite{Kapusta:2006pm}.
As in the case of the coupled $\phi^4$ theory, we here extend these rules to those for the first term in the parenthesis of \eqref{decomposition}.
We assume that the fermion field is the subsystem $A$ whose entanglement entropy is calculated and 
the photon field is 
the subsystem $B$ that is traced out. We can easily find the rules for the opposite case.
Momentum-space Feynman rules are 
\begin{figure}[H]
\vspace{0.3cm}
\begin{center}
   \includegraphics[clip, width=10.9cm]{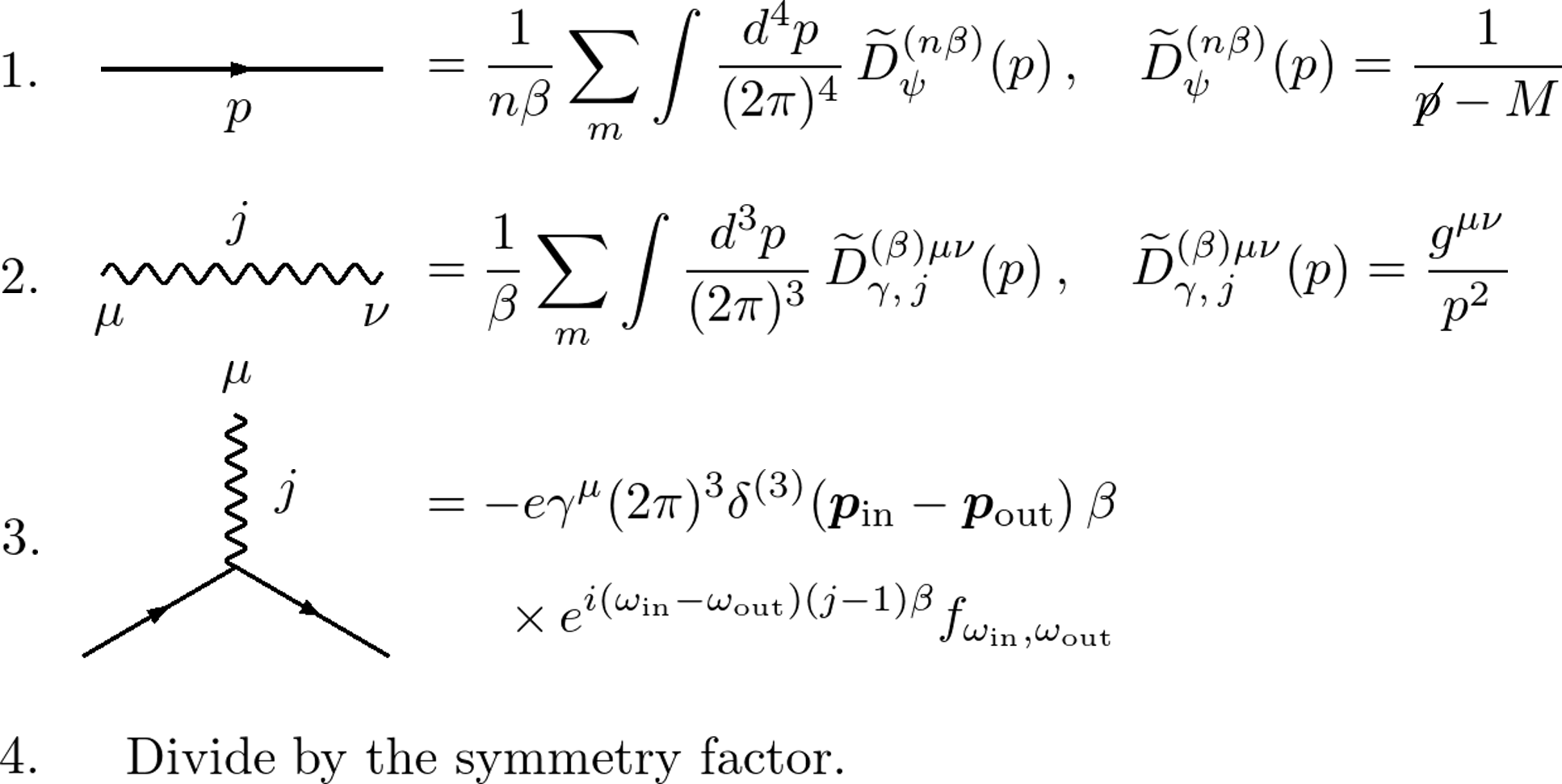}
\vspace{0cm}
\end{center}
\end{figure}
\noindent In addition to these rules, we need to multiply a factor of $-1$ for each fermion loop as usual.
The rule $1$ corresponds to a fermion line and $\widetilde{D}_\psi^{(n\beta)} (p)$ is the momentum-space propagator.
Here, $p^0 = i \widetilde{\omega}_m$ and $\widetilde{\omega}_m = 2\pi (m + \frac{1}{2}) T /n$, 
where a factor of $1/2$ comes from antiperiodicity of the fermion field.
The rule $2$ corresponds to a photon line for each $j$ ($j = 1, \cdots , n$).
For the photon propagator, $g^{\mu\nu} = (1, -1, -1, -1)$ is the metric and $p^0 = i \omega_m = i 2 \pi m T$.
The rule $3$ is for a QED vertex and energy conservation is not necessarily satisfied as discussed in section~\ref{Feynman}.

\begin{figure}[!t]
\vspace{-0.5cm}
\hspace{0.2cm}
 \begin{subfigure}{0.24\columnwidth}
\vspace{0.1cm}
\centering
   \includegraphics[clip, width=3.5cm]{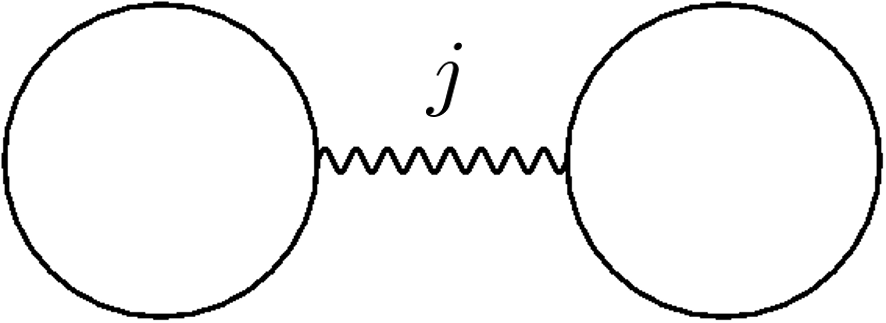}
\vspace{0.6cm}
\caption{}
\label{QED(a)}
\end{subfigure}
\vspace{-0.2cm}
\hspace{0.3cm}
\centering
 \begin{subfigure}{0.24\columnwidth}
\centering
   \includegraphics[clip, width=1.5cm]{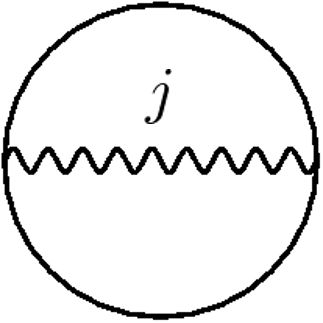}
\vspace{0.5cm}
\caption{}
\label{QED(b)}
\end{subfigure}
\hspace{-0.3cm}
 \begin{subfigure}{0.24\columnwidth}
\vspace{-0.1cm}
\centering
   \includegraphics[clip, width=1.8cm]{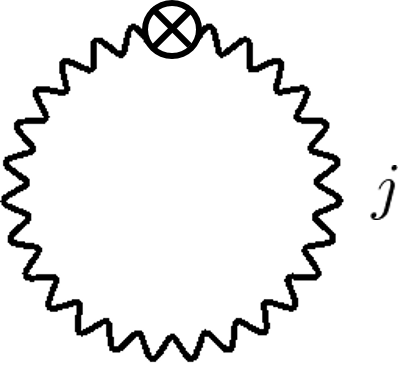}
\vspace{0.5cm}
\caption{}
\label{QED(d)}
\end{subfigure}
\hspace{-0.5cm}
 \begin{subfigure}{0.24\columnwidth}
\vspace{-0.1cm}
\centering
   \includegraphics[clip, width=1.5cm]{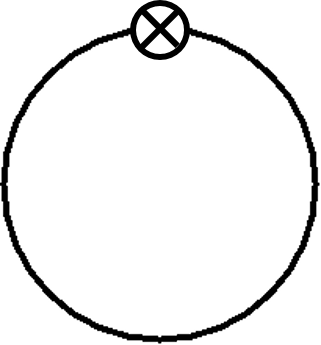}
\vspace{0.5cm}
\caption{}
\label{QED(c)}
\end{subfigure}
\vspace{0.5cm}
  \caption{Feynman diagrams which give the leading order correction in QED.
The last two diagrams come from the counterterms.}
  \label{fig:QEDdiagram}
\end{figure}

Figure~\ref{fig:QEDdiagram} shows Feynman diagrams which
contribute to the first term in the parenthesis of \eqref{decomposition} at the leading order.
The last two diagrams come from the counterterms that cancel divergence.
Let us first consider the diagram~\ref{QED(a)} which contains two fermion loops.
The contribution from this diagram actually vanishes in QED.
Using momentum-space Feynman rules, we can see that explicitly as 
\begin{equation}
\begin{split}
(\text{Fig.~\ref{QED(a)}}) &\propto \sum_{m_1} \int \frac{d^3 k_1}{(2 \pi)^3} \, 
\sum_{m_2} \int \frac{d^3 k_2}{(2 \pi)^3} 
\frac{ {\rm Tr} \left[ \gamma_\mu \left( \slashed{k}_1+M \right) \right] \times
{\rm Tr} \left[ \gamma^\mu \left( \slashed{k}_2+M \right) \right] }{\left( k_1^2 - M^2 \right) \left( k_2^2 - M^2 \right)} \\[1ex]
&\propto\sum_{m_1} \int \frac{d^3 k_1}{(2 \pi)^3}
\sum_{m_2} \int \frac{d^3 k_2}{(2 \pi)^3} 
\frac{- k_1^0 \cdot k_2^0 + \bm{k}_1 \cdot \bm{k}_2}{\left( k_1^2 - M^2 \right) \left( k_2^2 - M^2 \right)} = 0 \, ,
\end{split}
\end{equation}
where $k_1^0 = i \widetilde{\omega}_{m_1}$ and $k_2^0 = i \widetilde{\omega}_{m_2}$.
In the last equality, we have used the fact that the two terms in the integrand are odd under $k_1^0 \rightarrow - k_1^0$
and $\bm{k}_1 \rightarrow -\bm{k}_1$ respectively.

We next consider the diagram~\ref{QED(b)}.
The symmetry factor is $F=2$.
The Feynman rules lead to
\begin{equation}
\begin{split}
(\text{Fig.~\ref{QED(b)}}) = 
(-1) \, \cdot &\frac{1}{2} \cdot \sum_j \frac{1}{n \beta} \sum_{m_1} \int \frac{d^3 k_1}{(2 \pi)^3}
\frac{1}{n \beta} \sum_{m_2} \int \frac{d^3 k_2}{(2 \pi)^3} \frac{1}{\beta} \sum_{m_3} \int \frac{d^3 k_3}{(2 \pi)^3} \, \biggl\{ \\[1ex]
&(-e)^2 (2 \pi)^3 \delta^{(3)} (\bm{k}_1 - \bm{k}_2 - \bm{k}_3) \, \beta f_{\widetilde{\omega}_{m_1}, \widetilde{\omega}_{m_2} + \omega_{m_3}} 
\, e^{i (j-1) (\widetilde{\omega}_{m_1} - \widetilde{\omega}_{m_2})}\\[1.5ex]
\times \, &(2 \pi)^3 \delta^{(3)} (- \bm{k}_1 + \bm{k}_2 + \bm{k}_3) \,  \beta  f_{-\widetilde{\omega}_{m_1}, -\widetilde{\omega}_{m_2} - \omega_{m_3}} 
\, e^{-i (j-1) (\widetilde{\omega}_{m_1} - \widetilde{\omega}_{m_2})} \\[1.5ex]
\times \, &\frac{{\rm Tr} \left[ \gamma^\mu \left( \slashed{k}_1 + M \right)
\gamma_\mu \left( \slashed{k}_2 + M \right) \right]}{k_3^2 \left( k_1^2 - M^2 \right) \left( k_2^2 - M^2 \right)} \biggr\} \nonumber
\end{split}
\end{equation}
\begin{equation}
\begin{split}
= - \frac{1}{2} n &e^2 \beta V \frac{1}{n \beta} \sum_{m_1} \int \frac{d^3 k_1}{(2 \pi)^3}
\frac{1}{n \beta} \sum_{m_2} \int \frac{d^3 k_2}{(2 \pi)^3} \frac{1}{\beta} \sum_{m_3} \int \frac{d^3 k_3}{(2 \pi)^3} \, \biggl\{ \\[1ex]
&(2 \pi)^3 \delta^{(3)} (\bm{k}_1 - \bm{k}_2 - \bm{k}_3) \, \beta f_{\widetilde{\omega}_{m_1}, \widetilde{\omega}_{m_2} + \omega_{m_3}} 
f_{-\widetilde{\omega}_{m_1}, -\widetilde{\omega}_{m_2} - \omega_{m_3}} \\[1.5ex]
\times \, &\frac{8 \left( 2 M^2 - k_1 \cdot k_2 \right)}{k_3^2 \left( k_1^2 - M^2 \right) \left( k_2^2 - M^2 \right)} \biggr\} \, .
\end{split}
\end{equation}
Here,  $k_1^0 = i \widetilde{\omega}_{m_1}$ and $k_2^0 = i \widetilde{\omega}_{m_2}$ are the zeroth components of the fermion four momentum 
and $k_3^0 = i \omega_{m_3}$ is the zeroth component of the photon four momentum.
The minus sign comes from a fermion loop.
This diagram provides the first example that the factor $f$ is nontrivial.
There is a technical issue to calculate 
$f_{\widetilde{\omega}_{m_1}, \widetilde{\omega}_{m_2} + \omega_{m_3}}
f_{-\widetilde{\omega}_{m_1}, -\widetilde{\omega}_{m_2} - \omega_{m_3}}$, 
which 
is explained in Appendix~\ref{calculation}.
The sums of $m_1$, $m_2$ and $m_3$ can be performed by using the following relations for 
the fermion energy sum and photon energy sum: 
\begin{equation}
\begin{split}
&\frac{1}{n\beta} \sum_{m_1} \frac{1}{k_1^2 -M^2} \mathcal{I} (k_1^0, k_2^0 , k_3^0 )
= \frac{1}{2E_1} \mathcal{I} (E_1 , k_2^0, k_3^0) \, n_1 + \frac{1}{2E_1} \mathcal{I} (-E_1 , k_2^0, k_3^0) (n_1 -1) \, , \\[2ex]
&\frac{1}{\beta} \sum_{m_3} \frac{1}{k_3^2} \mathcal{I} (k_1^0, k_2^0 , k_3^0)
= -\frac{1}{2\omega} \mathcal{I} (k_1^0 , k_2^0, \omega) N - \frac{1}{2\omega} \mathcal{I} (k_1^0 , k_2^0, -\omega) (N +1) \, , \label{sumformula}
\end{split}
\end{equation}
where $\omega = |\bm{k}_3|$ and $E_1 = \sqrt{\bm{k}_1^2 + M^2}$, $E_2 = \sqrt{\bm{k}_2^2 + M^2}$ and $\mathcal{I} (k_1^0, k_2^0 , k_3^0 )$
is some function which has no singularities along with the imaginary axes.
The fermion and boson occupation numbers are
\begin{equation}
\begin{split}
n_1 = \frac{1}{e^{n\beta E_1} +1} \, , \qquad n_2 = \frac{1}{e^{n\beta E_2} +1} \, , \qquad N = \frac{1}{e^{\beta \omega} -1}  \, .
\end{split}
\end{equation}
Using the result of Appendix~\ref{calculation} for 
$f_{\widetilde{\omega}_{m_1}, \widetilde{\omega}_{m_2} + \omega_{m_3}}
f_{-\widetilde{\omega}_{m_1}, -\widetilde{\omega}_{m_2} - \omega_{m_3}}$ 
and the above relations, we obtain
\begin{equation}
\begin{split}
\frac{1}{2} n e^2 \beta V &\int \frac{d^3 k_1}{(2 \pi)^3}
\int \frac{d^3 k_2}{(2 \pi)^3}  \int \frac{d^3 k_3}{(2 \pi)^3} \left\{  (2 \pi)^3 \delta^{(3)} (\bm{k}_1 - \bm{k}_2 - \bm{k}_3)
\, \frac{2D(E_1, E_2, \omega)}{2E_1 \, 2E_2 \, 2 \omega} \right\} . \label{propton}
\end{split}
\end{equation}
Here, we have defined a function $D$ contributed from the diagram~\ref{QED(b)} as
\begin{equation}
\begin{split}
D (E_1, E_2, \omega) \bigr|^{\text{Fig.~\ref{QED(b)}}}  = \,\,&n_1 (1-n_2) (1+N) \mathcal{F}_{+++} + (1-n_1) n_2 N \mathcal{G}_{+++} \\[0.5ex]
&+n_1 (1-n_2) N \mathcal{F}_{++-} + (1-n_1) n_2 (1+N) \mathcal{G}_{++-} \\[1ex]
&-n_1 n_2 (1+N) \mathcal{F}_{+-+} - (1-n_1) (1-n_2) N \mathcal{G}_{+-+} \\[1ex]
&-n_1 n_2 N \mathcal{F}_{+--} - (1-n_1) (1-n_2) (1+N) \mathcal{G}_{+--}  \, . \label{Ddef}
\end{split}
\end{equation}
For definition of the functions $\mathcal{F}$, $\mathcal{G}$, see Appendix~\ref{calculation}.

When the first term in the parenthesis of \eqref{decomposition} is proportional to $n$ as in \eqref{propton},
it is easier to use the following expression of the entanglement entropy by rewriting \eqref{decomposition} with the limit of $n \rightarrow 1$:
\begin{equation}
\begin{split}
S_A  &=S_{A, 0}  - \frac{\partial}{\partial n} \left( \frac{1}{n} \log \frac{\widetilde{Z}_{\rm tot}^{(n\beta)}}{Z_{A,0}^{(n\beta)}
 ( Z_{B,0}^{(\beta)} )^n } \right) \Biggr|_{n =1}. \label{derivative}
\end{split}
\end{equation}
Inserting \eqref{propton} into this expression,
we obtain
\begin{equation}
\begin{split}
S_A - S_{A, 0} =\frac{1}{2} e^2 \beta V \int \frac{d^3 k_1}{(2 \pi)^3}
\int \frac{d^3 k_2}{(2 \pi)^3}  \int \frac{d^3 k_3}{(2 \pi)^3} \, \biggl\{  &(2 \pi)^3 \delta^{(3)} (\bm{k}_1 - \bm{k}_2 - \bm{k}_3) \\[1ex]
&\times \frac{2}{E_1 E_2 \, \omega} \left( - \frac{1}{8} \frac{\partial D}{\partial n} \biggr|_{n=1} \right) \biggr\} \, , \label{SAb}
\end{split}
\end{equation}
where
\begin{equation}
\begin{split}
- \frac{1}{8} \frac{\partial D}{\partial n} \biggr|_{n=1}^{\text{Fig.~\ref{QED(b)}}}&= - 2 \beta \omega \left[ \frac{M^2}{(E_1 - E_2)^2 - \omega^2} + \frac{M^2}{(E_1 + E_2)^2 - \omega^2} + 1 \right]  \\[1ex]
&\qquad \qquad \qquad \times \left[ E_1 n_1 (1-n_1) n_2 + E_2 n_1 n_2 (1-n_2) \right] \\[1ex]
&- 4 \beta E_1 E_2 \, n_1 (1-n_1) N \\[1ex]
&- 2 \beta E_1  n_1 (1 -n_1) \left[ E_2
- \omega + \frac{2M^2 (E_2 + \omega)}{(E_2 + \omega)^2 - E_1^2 } \right] \\[1.5ex]
&+\frac{\left(  2M^2 - E_1 E_2 + \bm{k}_1 \cdot \bm{k}_2 \right)\left(E_1 - E_2 \right)
}{\left(E_1 - E_2 + \omega \right)^2}
\left( n_1 N -n_2 N + n_1 n_2 -n_2 \right) \\[1ex]
&+\frac{\left(  2M^2 + E_1 E_2 + \bm{k}_1 \cdot \bm{k}_2 \right) \left(E_1 + E_2 \right)}{\left(E_1 + E_2 - \omega \right)^2}
\left( - n_1 n_2 - n_1 N -n_2 N + N \right) \\[1ex]
&+\frac{\left(  2M^2 + E_1 E_2 + \bm{k}_1 \cdot \bm{k}_2 \right) \left(E_1 + E_2 \right)
}{\left(E_1 + E_2 + \omega \right)^2} \\[1ex]
&\qquad \times \left( - n_1 N - n_2 N + n_1 n_2 -n_1 -n_2 + N+1\right) \\[1ex]
&+\frac{\left(  2M^2 - E_1 E_2 + \bm{k}_1 \cdot \bm{k}_2 \right)\left(E_1 - E_2 \right)}{\left(E_1 - E_2 - \omega \right)^2} 
\left( n_1 N -n_2 N  - n_1 n_2 + n_1 \right) \, . \label{Dprimeb2}
\end{split}
\end{equation}
In this expression, there are terms linear in $n_1$, $n_2$ or $N$ and the term without their dependence.
They are divergent terms when we perform all the integrals of $\bm{k}_1$, $\bm{k}_2$ and $\bm{k}_3$. 
The other terms are finite.
We will consider the meaning of divergence below, but some of the terms are canceled by the contributions
from the diagrams~\ref{QED(d)}, \ref{QED(c)} with the counterterms, as we will see next.

Let us evaluate the diagrams~\ref{QED(d)}, \ref{QED(c)}.
As in the case of the coupled $\phi^4$ theory, we can consider the usual
zero-temperature field theory and compute the sum of all 1PI insertions
into the photon and fermion propagators to determine the parameters in the counterterms, setting renormalization conditions.
Since the procedure is the same as that of the usual finite-temperature field theory except for complications with the index $j$,
we can use the results of Ref.~\cite{Kapusta:1979fh} for the contributions from the diagrams~\ref{QED(d)}, \ref{QED(c)} with the following modification.
When $\beta \mathcal{Z}_1 (\beta)$ and $\beta \mathcal{Z}_2 (\beta)$ denote the contributions from the corresponding diagrams
in the ordinary finite-temperature field theory to the diagrams~\ref{QED(d)} and \ref{QED(c)} respectively,
the contributions from the diagrams~\ref{QED(d)} and \ref{QED(c)} can be written as
\begin{equation}
\begin{split}
(\text{Fig.~\ref{QED(d)}}) = n \beta \mathcal{Z}_1 (\beta) \, , \qquad (\text{Fig.~\ref{QED(c)}}) = n \beta \mathcal{Z}_2 (n\beta) \, .
\end{split}
\end{equation}
The contribution from the diagram~\ref{QED(d)} is proportional to $n$ because of the $j$ index in the photon line.
Note that the contribution from the diagram~\ref{QED(c)} can be obtained
by the replacement $\beta \rightarrow n \beta$ in $\beta \mathcal{Z}_2 (\beta)$.
Then, the expressions for the correction terms in the entanglement entropy are given by
\begin{equation}
\begin{split}
&(S_A - S_{A, 0}) \Bigr|^{\text{Fig.~\ref{QED(d)}}} = \frac{1}{1-n}
\left[ n \beta \mathcal{Z}_1 (\beta) - n \beta \mathcal{Z}_1 (\beta) \right] \biggr|_{n=1} = 0 \, , \\[1.5ex]
&(S_A - S_{A, 0}) \Bigr|^{\text{Fig.~\ref{QED(c)}}} = \frac{1}{1-n}
\left[ n \beta \mathcal{Z}_2 (n\beta) - n \beta \mathcal{Z}_2 (\beta) \right] \biggr|_{n=1}
= - \beta \frac{\partial }{\partial n } \mathcal{Z}_2 (n\beta) \biggr|_{n=1} \, .
\end{split}
\end{equation}
There is no contribution to the entanglement entropy from the diagram~\ref{QED(d)}.
Using the result of Ref.~\cite{Kapusta:1979fh}, we can write down the contribution from the diagram~\ref{QED(c)} explicitly.
The expression is given by \eqref{SAb} with
\begin{equation}
\begin{split}
- \frac{1}{8} \frac{\partial D}{\partial n} \biggr|_{n=1}^{\text{Fig.~\ref{QED(c)}}}=
 2 \beta E_1  n_1 (1 -n_1) \left[ E_2
- \omega + \frac{2M^2 (E_2 + \omega)}{(E_2 + \omega)^2 - E_1^2 } \right] .
\end{split}
\end{equation}
Note that this has the same expression as the terms in the third line of \eqref{Dprimeb2} except for the sign
so that these two divergent contributions are canceled with each other.

We now discuss the other divergent terms in \eqref{Dprimeb2}.
In quantum field theory, a physical particle is dressed in clothing
\cite{Greenberg:1900zz}, a virtual cloud of particles.
Then, when we trace out the subsystem $B$, we should have taken account of this phenomenon
and taken the proper Hilbert space of the subsystem $B$ with physical one-particle states
to find the entanglement entropy of the subsystem composed of physical particles $A$.
In \eqref{Dprimeb2}, the terms linear in $n_1$, $n_2$ or $N$ and the term without their dependence
can be considered as remnants of the improper choice of the traced out Hilbert space and 
have the meaning of the entanglement entropy of the one-particle states and the vacuum respectively,
which are present even after the decoupling.
Since the entanglement entropy in which we are interested is the one between two physical particles,
we just remove these terms and consider only the finite terms.
Then, our final result of the contribution to the first term in the parenthesis of \eqref{decomposition} at the leading order
is given by \eqref{SAb} with
\begin{equation}
\begin{split}
- \frac{1}{8} \frac{\partial D}{\partial n} \biggr|_{n=1}&= - 2 \beta \omega \left[ \frac{M^2}{(E_1 - E_2)^2 - \omega^2} + \frac{M^2}{(E_1 + E_2)^2 - \omega^2} + 1 \right]  \\[1ex]
&\qquad \qquad \qquad \times \left[ E_1 n_1 (1-n_1) n_2 + E_2 n_1 n_2 (1-n_2) \right] \\[1ex]
&- 4 \beta E_1 E_2 \, n_1 (1-n_1) N \\[1ex]
&+\frac{\left(  2M^2 - E_1 E_2 + \bm{k}_1 \cdot \bm{k}_2 \right)\left(E_1 - E_2 \right)
}{\left(E_1 - E_2 + \omega \right)^2}
\left( n_1 N -n_2 N + n_1 n_2 \right) \\[1ex]
&+\frac{\left(  2M^2 + E_1 E_2 + \bm{k}_1 \cdot \bm{k}_2 \right) \left(E_1 + E_2 \right)}{\left(E_1 + E_2 - \omega \right)^2}
\left( - n_1 n_2 - n_1 N -n_2 N \right) \\[1ex]
&+\frac{\left(  2M^2 + E_1 E_2 + \bm{k}_1 \cdot \bm{k}_2 \right) \left(E_1 + E_2 \right)
}{\left(E_1 + E_2 + \omega \right)^2} \left( - n_1 N - n_2 N + n_1 n_2 \right) \\[1ex]
&+\frac{\left(  2M^2 - E_1 E_2 + \bm{k}_1 \cdot \bm{k}_2 \right)\left(E_1 - E_2 \right)}{\left(E_1 - E_2 - \omega \right)^2} 
\left( n_1 N -n_2 N  - n_1 n_2 \right) \, . \label{final}
\end{split}
\end{equation}
Infrared behavior of this result is discussed in Appendix~\ref{infra}.
Figure~\ref{fig:resultQED} shows 
the size of the correction $\Delta S_A \equiv S_A - S_{A,0}$ as a function of $M /T$ in QED.
Here, $\alpha \equiv e^2 / (4 \pi)$.
The solid (dashed) curve represents that $\Delta S_A$ is positive (negative).
We can see that $\Delta S_A$ changes its sign depending on the fermion mass $M$.

\begin{figure}[!t]
%\vspace{-1.5cm}
  \begin{center}
   \includegraphics[clip, width=7.5cm]{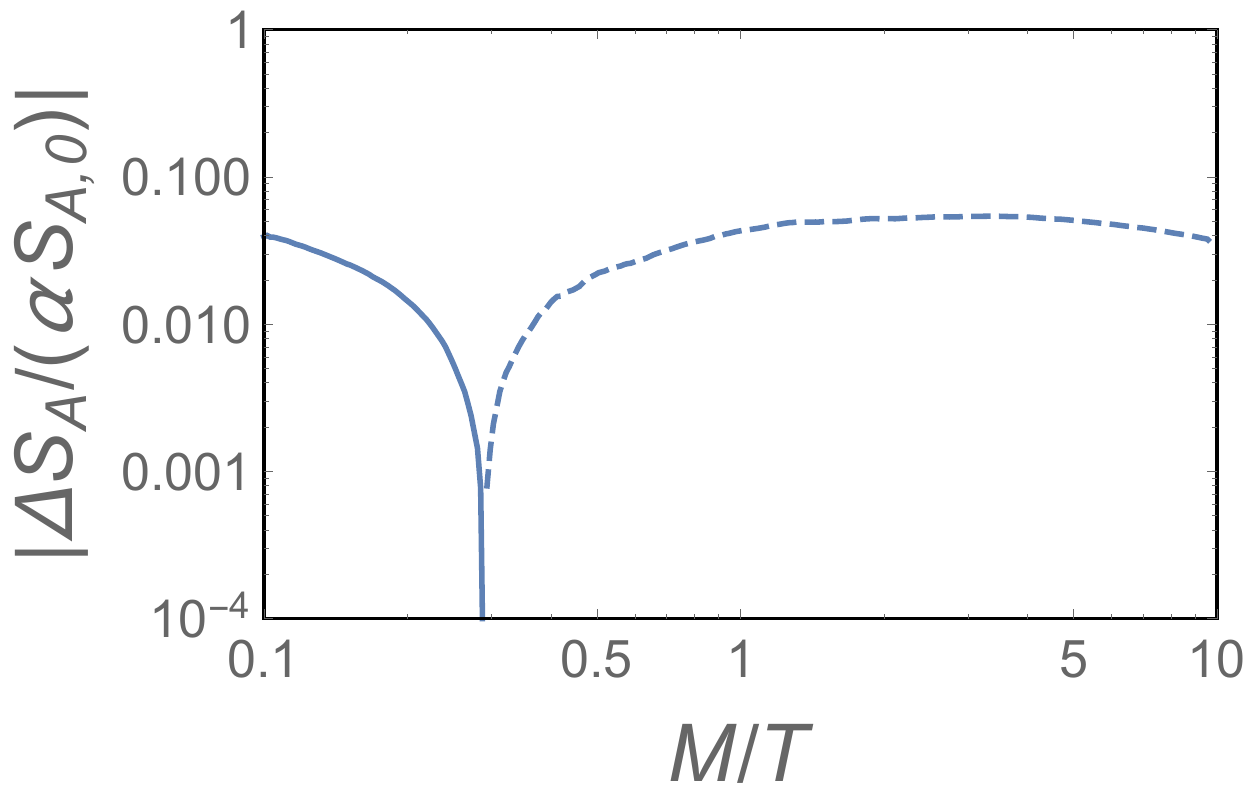}
  \end{center}
%\vspace{-0.8cm}
  \caption{The size of the correction $\Delta S_A \equiv S_A - S_{A,0}$ as a function of $M /T$ in QED.
Here, $\alpha \equiv e^2 / (4 \pi)$.
  The solid (dashed) curve represents that $\Delta S_A$ is positive (negative).
  }
  \label{fig:resultQED}
\end{figure}

\section{Yukawa theory}\label{Yukawa}

The third model we consider in this paper is the Yukawa theory of a scalar-fermion system.
We evaluate the entanglement entropy of a fermion subsystem in this theory, tracing out the scalar field.
The calculation is similar to that of QED, 
though there are some differences. 
The interaction Lagrangian of the Yukawa theory in $3+1$ dimensions is presented as
\begin{equation}
\begin{split}
\mathcal{L}_I (\psi, \phi)  &= g \phi \bar{\psi} \psi   \, .
\end{split}
\end{equation}
Here, $g$ is a coupling constant. Since there is no symmetry such as $\phi \rightarrow - \phi$ in this theory, the counterterm linear in $\phi$
should be included: 
\begin{equation}
\begin{split}
\mathcal{L}_{\rm counter} (\psi, \phi)  &\supset \delta_\phi \phi   \, . \label{tadpole}
\end{split}
\end{equation}
As in the case of QED, we assume that the chemical potential of the fermion field is zero for simplicity.
The fermion field is the subsystem $A$ whose entanglement entropy is calculated and the traced out subsystem $B$ is
the scalar field.
Then, the momentum-space vertex rule for the first term in the parenthesis of \eqref{decomposition} is
\begin{figure}[H]
\vspace{0.0cm}
\begin{center}
   \includegraphics[clip, width=7.5cm]{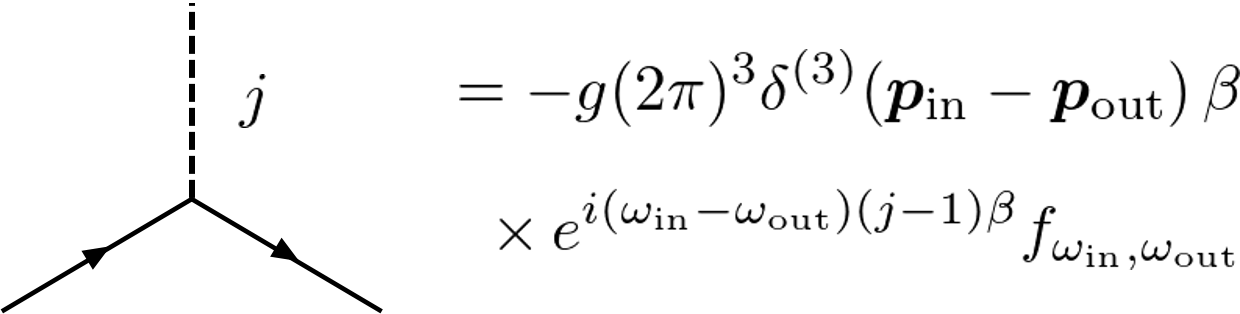}
\vspace{-0.3cm}
\end{center}
\end{figure}
\noindent The other rules are the same as those of the coupled $\phi^4$ theory and QED.

\begin{figure}[!t]
\vspace{-0.5cm}
\hspace{0.7cm}
 \begin{subfigure}{0.3\columnwidth}
\vspace{0.15cm}
\centering
   \includegraphics[clip, width=3.5cm]{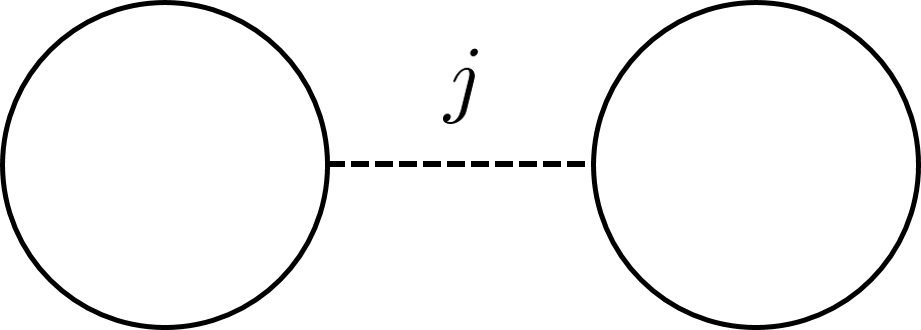}
\vspace{0.3cm}
\caption{}
\label{Yukawa(a)}
\end{subfigure}
\vspace{-0.2cm}
\hspace{0cm}
\centering
 \begin{subfigure}{0.3\columnwidth}
\centering
   \includegraphics[clip, width=2.5cm]{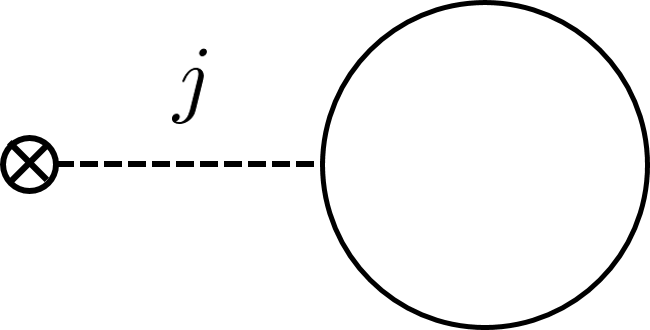}
\vspace{0.3cm}
\caption{}
\label{Yukawa(b)}
\end{subfigure}
\vspace{-0.2cm}
\hspace{-0.5cm}
\centering
 \begin{subfigure}{0.3\columnwidth}
\vspace{0.15cm}
\centering
   \includegraphics[clip, width=1.5cm]{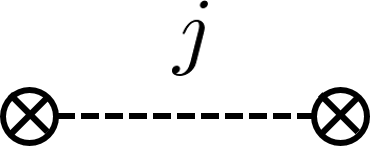}
\vspace{0.8cm}
\caption{}
\label{Yukawa(b)}
\end{subfigure}
\vspace{0.5cm}
  \caption{A part of Feynman diagrams which gives the leading order correction in the Yukawa theory.
The last two diagrams come from the counterterm. The symmetry factor is $(a) \, F=8$, $(b) \, F=2$, $(c) \, F=2$. }
  \label{fig:Yukawadiagram}
\end{figure}

The quantum correction starts from two-loop diagrams.
Figure~\ref{fig:Yukawadiagram} shows a part of Feynman diagrams which contribute to
the first term in the parenthesis of \eqref{decomposition} at the leading order.
The right two diagrams come from the counterterm \eqref{tadpole}.
There are also diagrams of \ref{QED(b)}, \ref{QED(d)}, \ref{QED(c)} with a photon line replaced by a scalar line.

Let us first consider the diagram~\ref{Yukawa(a)} which contains two fermion loops.
The symmetry factor is $F =8$.
Unlike the similar diagram in QED, this diagram does not vanish in the Yukawa theory.
Using momentum-space Feynman rules, we obtain
\begin{equation}
\begin{split}
(\text{Fig.~\ref{Yukawa(a)}} ) = \frac{1}{8} n &g^2 \beta V \frac{16M_\psi^2}{M_\phi^2} \left( \frac{1}{n \beta}
\sum_{m} \int \frac{d^3 k}{(2 \pi)^3} \, 
\frac{1}{  k^2 - M_\psi^2 } \right)^2 .
\end{split}
\end{equation}
Here,  $k^0 = i \widetilde{\omega}_{m}$ is the zeroth component of the fermion field and
$M_\psi, M_\phi$ are the fermion mass and the scalar mass respectively.
The sum of $m$ is performed by using the formula \eqref{sumformula} for a fermion field.
Then, we find
\begin{equation}
\begin{split}
\frac{1}{8} n &g^2 \beta V \frac{16M_\psi^2}{M_\phi^2} \left[ \int \frac{d^3 k_1}{(2 \pi)^3} 
\left( \frac{n_1 }{  E_1 } - \frac{1}{ 2 E_1 } \right) \right]^2 ,
\label{glasses}
\end{split}
\end{equation}
where $E_1 = \sqrt{\bm{k}_1^2 + M_\psi^2}$ and $n_1 = \left( e^{n \beta E_1} +1 \right)^{-1}$ is the fermion occupation number.
In this expression, there are a term linear in $n_1$ and a term without its dependence.
They are divergent when we perform all the integrals.
These divergent terms are canceled by the contributions from the last two diagrams of Figure~\ref{fig:Yukawadiagram}
with the appropriate choice of the parameter $\delta_\phi$
in the same way as we have done in the coupled $\phi^4$ theory or QED
(we have to be careful about the symmetry factor of each diagram, as presented in Figure~\ref{fig:Yukawadiagram}).

The contributions from the diagrams of \ref{QED(b)}, \ref{QED(d)}, \ref{QED(c)} with a photon line replaced by a scalar line
are calculated in the same way as those of QED.
The final expression of the entanglement entropy is given by
\begin{equation}
\begin{split}
S_A - S_{A, 0} &= \frac{1}{4}  g^2 \beta V \frac{16M_\psi^2}{M_\phi^2} \int \frac{d^3 k_1}{(2 \pi)^3} \int \frac{d^3 k_2}{(2 \pi)^3} \,
\frac{\beta }{  E_1 } \, n_1 n_2 (1 -n_2) \\[1.5ex]
&\quad+\frac{1}{2} g^2 \beta V \int \frac{d^3 k_1}{(2 \pi)^3}
\int \frac{d^3 k_2}{(2 \pi)^3}  \int \frac{d^3 k_3}{(2 \pi)^3} \biggl\{  (2 \pi)^3 \delta^{(3)} (\bm{k}_1 - \bm{k}_2 - \bm{k}_3) \\[1ex]
&\qquad \qquad \qquad \qquad \qquad \qquad \qquad \qquad \quad
\times \frac{2}{E_1 E_2 E_3} \left( - \frac{1}{8} \frac{\partial D}{\partial n} \biggr|_{n=1} \right) \biggr\} \, .
\end{split}
\end{equation}
Here, $E_3 = \sqrt{\bm{k}_3^2 + M_\phi^2}$ is the scalar energy, $N = \left( e^{\beta E_3} - 1 \right)^{-1}$
is the boson occupation number and
$- \frac{1}{8} \frac{\partial D}{\partial n} \biggr|_{n=1}$ is given in \eqref{final} with the replacement,
\begin{equation}
\begin{split}
\left(  2M^2 \mp E_1 E_2 + \bm{k}_1 \cdot \bm{k}_2 \right)
\rightarrow \frac{1}{2}\left(  M_\psi^2 \pm E_1 E_2 - \bm{k}_1 \cdot \bm{k}_2 \right) .
\end{split}
\end{equation}
For definition of the functions $\mathcal{F}$, $\mathcal{G}$ in the Yukawa theory, see Appendix~\ref{calculation}.
As in the case of QED, the divergent terms 
should be 
removed in the above result of the entanglement entropy.

\begin{figure}[!t]
%\vspace{-1.5cm}
  \begin{center}
   \includegraphics[clip, width=7cm]{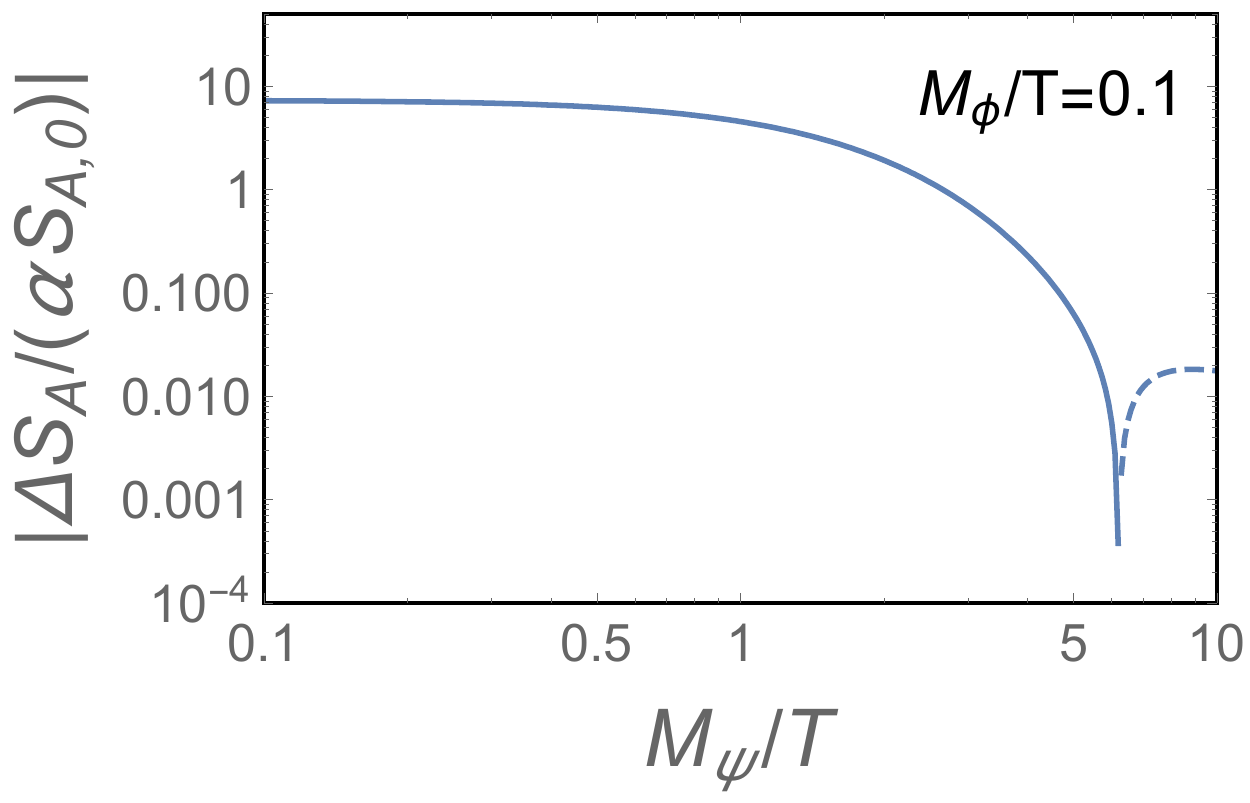}
   \qquad
      \includegraphics[clip, width=7cm]{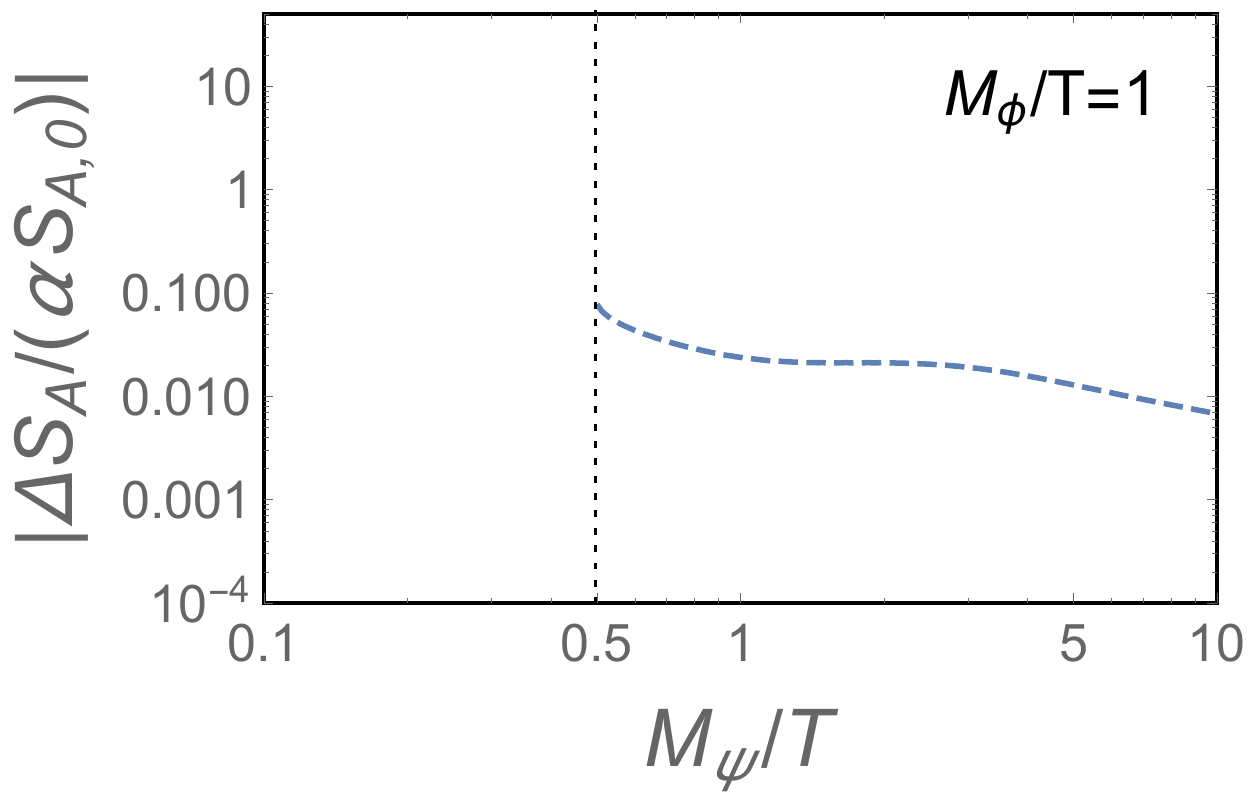}
  \end{center}
%\vspace{-0.8cm}
  \caption{The size of the correction $\Delta S_A$ as a function of $M_\psi /T$ in the Yukawa theory.
Here, $\alpha \equiv g^2 / (4 \pi)$. 
  The solid (dashed) curve represents that $\Delta S_A$ is positive (negative).
  We assume $M_\phi /T = 0.1$ and $M_\phi / T = 1$ in the left and right panels respectively. 
  }
  \label{fig:resultYukawa}
\end{figure}

Figure~\ref{fig:resultYukawa}
shows the size of the correction $\Delta S_A$ as a function of $M_\psi /T$ in the Yukawa theory.
Here, $\alpha \equiv g^2 / (4 \pi)$. 
The solid (dashed) curve represents that $\Delta S_A$ is positive (negative).
We assume $M_\phi /T = 0.1$ and $M_\phi / T = 1$ in the left and right panels respectively. 
Since the contribution from the diagram~\ref{Yukawa(a)}
diverges in the limit of $M_\phi / M_\psi  \rightarrow 0$ and $M_\psi / T \rightarrow 0$, 
one might be worried about the breakdown of perturbation theory. 
(See the left panel.
In the region of $M_\psi / T < 1$, values of $|\Delta S_A / \alpha S_{A,0}|$ are around $10$,
which is problematic for $\alpha \gtrsim 0.1$.)
However, as is well known in finite-temperature field theory,
the scalar field $\phi$ actually obtains a thermal mass in the presence of interactions.
This implies that 
we cannot take the limit of $M_\phi / M_\psi  \rightarrow 0$ and $M_\psi / T \rightarrow 0$
in our expression of the entanglement entropy
and perturbative expansion should be still valid even in this regime.
Mass resummation requires the next-to-leading order calculations, which are beyond the scope of the present paper,
and 
will be discussed in \cite{Future}.
In the right panel of the figure, there is a cutoff at $M_\psi /T = 0.5$
and the correction $\Delta S_A$ diverges for a smaller $M_\psi /T$.
This divergence comes from the removal of the terms linear in $N$ in order to ignore the entanglement entropy of one-particle states.
Since the scalar $\phi$ can decay into a pair of fermions and is unstable for $M_\psi < M_\phi /2$, 
the procedure to remove the terms linear in $N$ does not work in this region.

\section{Cosmological implications}
\label{cosmology}

In this section, we discuss cosmological implications of the entanglement entropy.
When instantaneous decoupling occurs, 
the thermodynamic entropy is no longer an appropriate fiducial quantity to 
describe a subsystem and should be replaced to the entanglement entropy.
We investigate the possible effect of quantum entanglement on dark radiation and dark matter
and also present a concrete scenario of instantaneous decoupling.

Here, let us comment on the time when the entanglement entropy is evaluated.
Since the entanglement entropy derives from the correlation between two particles $A, B$,
it is reasonable to adopt the time of the last scattering for that time.
The time of the last scattering $t_{\rm LS}$ is defined as
\begin{equation}
\begin{split}
 \int_{t_{\rm LS}}^{t_p} \Gamma_A (t) d t = 1 \, , 
\end{split}
\end{equation}
where $t_p$ is the present time. 
The interaction rate at the time $t_{\rm LS}$ has to be distinguished from the rate at the time of decoupling $t_0$
because the interaction rate $\Gamma_A (t)$ changes rapidly during the time interval, $t_{\rm LS} \leq t \leq t_0$,
in the case of instantaneous decoupling.\footnote{
Consider a system of a relativistic particle $A$ interacting with a massive particle $B$.
As the number density of the particle $B$ drops exponentially below $T \sim M_B$ ($M_B$ is the mass of $B$),
the interaction rate becomes smaller than the expansion rate quickly, which is apparently similar to instantaneous decoupling.
However, the Boltzmann factor $e^{-M_B/T}$ does not make the change of the interaction rate fast enough
during $t_{\rm LS} \leq t \leq t_0$.
In this case, the entanglement entropy $S_A$ can be approximately estimated at the time of decoupling.
The quantum entangle effect in $S_A$ is typically accompanied with the same Boltzmann factor and exponentially small.}

\subsection{Dark radiation and dark matter}

Dark components of the Universe are now explored by precise measurements.
The energy density of dark radiation 
is conveniently expressed in terms of the effective number of neutrinos.
The present constraint on this number 
is~\cite{Ade:2015xua}
\begin{equation}
 N_{\rm eff} = 3.15 \pm 0.23 \, , \label{darkconst}
\end{equation}
which is consistent with the Standard Model (SM) prediction, $N_{\rm eff}^{(\rm SM)} = 3.046$. 
The ground-based Stage-IV cosmic microwave background (CMB) polarization experiment 
CMB-S4 measures $N_{\rm eff}$ 
with a precision of $\Delta N_{\rm eff} = 0.0156$ 
within $1 \sigma$ level~\cite{Wu:2014hta} (see also Ref.~\cite{Abazajian:2013oma}). 
When we consider a cosmological scenario with some dark radiation components,
it is reasonable to estimate a possible correction to $N_{\rm eff}$ from quantum entanglement,
as we now describe.

Suppose that a dark radiation $A$ or its mother particle, which decays into $A$, is decoupled from the SM sector
due to an instantaneous suppression of the interaction rate of $A$ with the SM sector.
After the decoupling, the entanglement entropy \eqref{entanglemententropy} of the subsystem $A$ is conserved
and a good fiducial quantity.
We can calculate the entanglement entropy by using the technique developed in the previous sections as long as
perturbative expansion is valid.
We assume that the self-interaction of the dark radiation $A$ becomes efficient or the mother particle decays
into $A$ after the decoupling
so that the subsystem enters thermal equilibrium again.
Then, the entanglement entropy should be expressed by 
the usual thermodynamic entropy, 
which enables us to define the temperature of the subsystem $A$.
If the self-interaction of $A$ is sufficiently weak,
the temperature of the subsystem $A$ can be calculated as
\begin{equation}
\begin{split}
T_A = \left( \frac{45}{2 \pi^2 g_A} \frac{S_A}{V} \right)^{1/3} ,
\end{split}
\end{equation}
where $g_A$ is the number of degrees of freedom of the dark radiation $A$
(there is an additional factor $7/8$ for a fermion). 
This temperature $T_A$ is different from $T_{A, 0}$ calculated by the naive application of the usual thermodynamics.
The energy density of the dark radiation $\rho_A$ can be calculated from its temperature
and is different from the naive estimation by the factor $\left( T_A / T_{A,0} \right)^4$.
Then, we obtain a correction to the effective number of neutrinos from the dark radiation including the entanglement effect,
\begin{equation}
\Delta N_{\rm eff} \equiv \rho_A \left( 2 \cdot \frac{7}{8} \cdot \frac{\pi^2}{30} \cdot T_\nu^4 \right)^{-1}
= \frac{8}{7} \frac{g_{A}}{2} \left( \frac{g_* (T_{D})}{43/4} \right)^{-4/3}
\left( \frac{T_A}{T_{A, 0}} \right)^4 , 
\end{equation}
where $T_\nu$ is the neutrino temperature and $g_* (T_{D})$ is the effective number of relativistic degrees of freedom 
at the decoupling temperature $T_{D}$. 
For example, 
$g_* = 3.36$ (10.75) for $T \ll 1 \ {\rm MeV}$ ($1 \ {\rm MeV} \lesssim T \lesssim 100 \ {\rm MeV}$) 
in the Standard Model. 
Although the effect of quantum  entanglement should not be larger than the classical contribution,
it can be relevant for the constraint \eqref{darkconst} and the future CMB-S4 experiment.

In addition to the energy density of dark radiation, the energy density of dark matter has been measured precisely,
$\Omega_{\rm DM} h^2 = 0.1186 \pm 0.0020$
\cite{Ade:2015xua}.
Although we do not explicitly describe a scenario where the entanglement effect is relevant for the constraint over again,
this measurement can be also an interesting channel to observe a correction from quantum entanglement.

\subsection{A scenario of instantaneous decoupling}

We now present a cosmological scenario of instantaneous decoupling. 
For the purpose of illustration,
we concentrate on the coupled $\phi^4$ theory whose self-interactions are turned off, $\lambda_A = \lambda_B = 0$.
The similar discussion may be possible for QED or the Yukawa theory.
Let us
first assume that the masses of particles $\phi_A$ and $\phi_B$ are much smaller than
the temperature before decoupling. 
Thermal equilibrium between $\phi_A$ and $\phi_B$ can be maintained by the scattering of $\phi_A$ and $\phi_B$ such as
\begin{equation}
\begin{split}
\phi_A \phi_B \leftrightarrow \phi_A \phi_B \, .
\end{split}
\end{equation}
The scattering rate of a particle $\phi_A$ with particles $\phi_B$ per unit time 
is given by 
\begin{equation}
\Gamma_A
= n_B \langle \sigma_{\phi_A \phi_B \rightarrow \phi_A \phi_B} v \rangle \, , \label{intrate}
\end{equation}
where $n_B = \frac{\zeta(3)}{\pi^2} T^3$ is the number density of $\phi_B$ and
$v = \left( \frac{1}{E_A} + \frac{1}{E_B} \right) p_\ast$
is the relative velocity of the two incoming particles.
Here, we assume $E_A \approx E_B \approx p_\ast$ and $p_\ast$ is the center-of-mass momentum.
The cross section is 
\begin{equation}
\begin{split}
\sigma_{\phi_A \phi_B \rightarrow \phi_A \phi_B} = \frac{\lambda^2}{16 \pi s} \, ,
\end{split}
\end{equation}
with $\sqrt{s} = E_A + E_B$ in the center-of-mass frame.
Thermal average of $\sigma_{\phi_A \phi_B \rightarrow \phi_A \phi_B} v$ is defined as
\begin{equation}
\begin{split}
\langle \sigma_{\phi_A \phi_B \rightarrow \phi_A \phi_B} v \rangle =
\frac{\int_0^\infty d p_\ast \, p_\ast^2 \, e^{- \sqrt{s}/T}
\sigma_{\phi_A \phi_B \rightarrow \phi_A \phi_B} v}{\int_0^\infty d p_\ast \, p_\ast^2 \, e^{- \sqrt{s} /T}} 
\approx \frac{\lambda^2}{16 \pi T^2} \, .
\end{split}
\end{equation}
To see if the decoupling of $\phi_A$ and $\phi_B$ occurs or not, we compare the interaction rate \eqref{intrate} with the Hubble expansion rate $H$.
In the radiation-dominated universe, the expansion rate is given by $H = \sqrt{\frac{4 \pi^3 G_N g_{\ast,T}}{45}} \, T^2$ where $G_N$ is the
Newton gravitational constant and $g_{\ast,T}$ is the total number of degrees of freedom of relativistic particles.
In the present setup, the interaction rate is proportional to $T$ while the expansion rate is proportional to $T^2$.
Then, with a sufficiently large $\lambda$, $\phi_A$ and $\phi_B$ are in thermal equilibrium as $T$ drops.

To realize the decoupling of $\phi_A$ and $\phi_B$, we make the field $\phi_B$ massive dynamically and
consider the decay of $\phi_B$ into newly introduced particles $\psi_C$. 
The mass generation of the field $\phi_B$ can be provided by a coupling of $\phi_B$ with a new scalar field $X$
which gets a vacuum expectation value as in the case of
the mass generation of the SM fermions by the Higgs field.
We assume $\psi_C$ is a Dirac fermion with mass $M$ 
and specifically consider the Lagrangian,
\begin{equation}
\begin{split}
\mathcal{L} = \frac{\lambda}{4} \phi_A^2 \phi_B^2 +  \frac{\kappa}{2} X^2 \phi_B^2
+ y \phi_B \bar{\psi}_C \psi_C + M \bar{\psi}_C \psi_C \, ,
\end{split}
\end{equation}
where $\kappa$ in the second term is a coupling constant and
the third term denotes the Yukawa interaction with the coupling constant $y$ which induces the decay of $\phi_B$.
When the scalar $X$ gets a vacuum expectation value (VEV) $\langle X \rangle = v_X$ at some temperature $T_v$,
the mass of the field $\phi_B$ changes into
$M'^2_B = \kappa v_X^2$.
If we take $M'_B \gg M$ and the coupling $y$ is sufficiently large, the decay $\phi_B \rightarrow \psi_C \bar{\psi}_C$ proceeds promptly
while $\phi_B$ cannot decay into $\phi_A$ because of their interaction form respecting the parity.
The decay rate of $\phi_B$ is given by
\begin{equation}
\begin{split}
\Gamma_{\phi_B \rightarrow \psi_C \bar{\psi}_C} = \frac{y^2}{16 \pi } M'_B \, .
\end{split}
\end{equation}
Here, we have ignored the mass $M$.
If this decay rate is much larger than the expansion rate, $\Gamma_{\phi_B \rightarrow \psi_C \bar{\psi}_C} \gg H$,
the instantaneous decoupling between $\phi_A$ and $\phi_B$ is realized
because the number density of $\phi_B$ drops suddenly and $\Gamma_{\phi_A \phi_B \rightarrow \phi_A \phi_B}  \ll H$.
The entanglement entropy of the subsystem of $\phi_A$ in the present scenario is given by \eqref{phi4result} with $\lambda_A = 0$
at two-loop level.
Since the time when the entanglement entropy is evaluated is
just before the generation of the VEV $v_X$, the field $\phi_B$ is still (almost) massless and its number density
is not suppressed by the Boltzmann factor, which gives a sizable quantum correction in the entanglement entropy.

\section{Conclusion}\label{conclusion}

In this paper, we have formulated the perturbation theory to derive the entanglement entropy of coupled quantum fields
and presented Feynman rules in the diagrammatic calculations.
Since it is not easy to evaluate the trace of $\rho_A \log \rho_A$ directly,
we first calculated the R\'{e}nyi entropy \eqref{Renyi} and took the limit of $n \rightarrow 1$.
We thus needed to evaluate the trace of a multiple product of the density matrix $\rho_A$.
Accordingly, a part of contributions to the R\'{e}nyi entropy at a given order of perturbative expansion
is described by diagrams with a new index $j  \, (= 1, \cdots , n)$ or the inverse temperature $n \beta$.
Using the developed rules, the entanglement entropy has been calculated in
an interacting scalar-scalar system (the coupled $\phi^4$ theory), QED and the Yukawa theory.
We have also discussed cosmological implications of the entanglement entropy.
The correction from quantum entanglement is relevant in circumstances of instantaneous decoupling.
We analyzed its possible effect on dark radiation and dark matter because
the measurements of their energy densities are now becoming more and more precise.
Finally, a concrete scenario of instantaneous decoupling was presented.

We now comment on possible directions of future investigation.
As discussed in section~\ref{Yukawa},
the thermal mass of an interacting field is generated at finite temperature.
The next-to-leading order calculations to find the thermal mass will be discussed in \cite{Future}.
In addition, it might be interesting to investigate other scenarios of instantaneous decoupling or some other circumstances
where the correction from quantum entanglement is relevant.

\acknowledgments

We are grateful to Tadashi Takayanagi for useful comments and discussions.
N.S. also would like to thank Pawel Caputa, Nilay Kundu, Masamichi Miyaji, and Kento Watanabe for useful discussions.
Y.N. is grateful to Pouya Asadi, Tom Banks and David Shih for useful comments and discussions after the first version came out.
Y.N. would like to thank the Department of Energy grant DE-SC0013607 for supporting his research in Harvard University.
N.S. is supported by Grant-in-Aid for the JSPS Fellowship No.15J02740.

\appendix

\section{The vertex factor calculation}\label{calculation}

We here present the calculation of the vertex factor $f_{\widetilde{\omega}_{m_1},  \widetilde{\omega}_{m_2} + \omega_{m_3}} \,
f_{-\widetilde{\omega}_{m_1},  -\widetilde{\omega}_{m_2} - \omega_{m_3}}$
which we encounter in the calculation of the diagram~\ref{QED(b)}.
From the definition of the factor $f$ in \eqref{fl}, we obtain
\begin{equation}
\begin{split}
&f_{\widetilde{\omega}_{m_1},  \widetilde{\omega}_{m_2} + \omega_{m_3}} \,
f_{-\widetilde{\omega}_{m_1},  -\widetilde{\omega}_{m_2} - \omega_{m_3}} \\[1ex]
&= \frac{1}{\beta^2} \int_0^\beta d \tau_1 \int_0^\beta d \tau_2 \, e^{i (\widetilde{\omega}_{m_1} - \widetilde{\omega}_{m_2} - \omega_{m_3})
(\tau_1 - \tau_2)}
\\[1ex]
&= \frac{1}{\beta^2} \int_0^\beta d \tau_1\int_0^{\tau_1} d \tau_2 \left( e^{i (\widetilde{\omega}_{m_1} - \widetilde{\omega}_{m_2} - \omega_{m_3})
(\tau_1 - \tau_2)} +
e^{-i (\widetilde{\omega}_{m_1} - \widetilde{\omega}_{m_2} - \omega_{m_3}) (\tau_1 - \tau_2)} \right) \, , \label{ffcalc}
\end{split}
\end{equation}
where we change the integration variables 
in such a way that $0 < \tau_1 - \tau_2 < \beta$ for later convenience. 

When we perform the sums of $m_1$, $m_2$ and $m_3$ by using the relations \eqref{sumformula} in the main text,
there is one caveat.
The relations of \eqref{sumformula} come from
\cite{Kapusta:2006pm}
\begin{equation}
\begin{split}
\frac{1}{n\beta} \sum_{m = - \infty}^{\infty} \mathcal{F} (p^0 = i \widetilde{\omega}_m)
= & \, -\frac{1}{2 \pi i} \int_{- i \infty + \epsilon}^{i \infty + \epsilon} dp^0 \,
\mathcal{F} (p^0) \frac{1}{e^{n\beta p^0}+ 1} \\[1ex]
&-\frac{1}{2 \pi i} \int_{- i \infty - \epsilon}^{i \infty - \epsilon} dp^0 \,
\mathcal{F} (p^0) \frac{1}{e^{-n\beta p^0} +1} \\[1ex]
& \, +\frac{1}{2 \pi i} \int_{- i \infty}^{i \infty} dp^0 \,
\mathcal{F} (p^0)   \, , 
\end{split}
\label{sumformula2}
\end{equation}
for a fermion and \eqref{Fformula} for a boson field.
$\mathcal{F} (p^0)$ is some function which has no singularities along the imaginary $p^0$ axis.
In the above relation for a fermion field,
the integral of the first term is performed by extending the contour to a closed contour going along the positive infinity
while the integrals of the second and third terms are performed by extending the contours to closed contours going along the negative infinity.
In the same way, we can perform the integrals in the relation of \eqref{Fformula}.
Thus 
we need to insert $1 = -{\rm exp} [ i n \beta \widetilde{\omega}_{m_2}  + i \beta \omega_{m_3}]$ 
or $1 = - {\rm exp} [ i n \beta \widetilde{\omega}_{m_1}]$ in the integrands of \eqref{ffcalc} 
so that the integrand falls off exponentially at $|p^0| \to \infty$ for each contour
after analytic continuation. 
Then we perform the integrals and obtain 
\begin{equation}
\begin{split}ｔ
&\frac{1}{\beta^2} \int_0^\beta d \tau_1\int_0^{\tau_1} d \tau_2 \, \Bigl( e^{i (\widetilde{\omega}_{m_1} - \widetilde{\omega}_{m_2} - \omega_{m_3}) 
(\tau_1 - \tau_2)}
\left( - e^{i n \beta \widetilde{\omega}_{m_2}  + i \beta \omega_{m_3}} \right) \\[1ex]
&\qquad \qquad \qquad \qquad \qquad \quad + e^{-i (\widetilde{\omega}_{m_1} - \widetilde{\omega}_{m_2} - \omega_{m_3}) (\tau_1 - \tau_2)}
\left( - e^{i n \beta \widetilde{\omega}_{m_1}} \right) \Bigr) \\[1ex]
&= \widetilde{\mathcal{F}} e^{n \beta k_2^0 + \beta k_3^0} + \widetilde{\mathcal{G}} \, e^{n \beta k_1^0} \, ,
\end{split}
\end{equation}
where
\begin{equation}
\begin{split}
&\widetilde{\mathcal{F}} (k_1^0, k_2^0, k_3^0) \equiv \frac{1}{\beta (k_1^0 - k_2^0 - k_3^0)} + \frac{1 - e^{\beta (k_1^0 - k_2^0 - k_3^0) }}{\beta^2 (k_1^0 - k_2^0 - k_3^0 )^2}  \, , \\[1.5ex]
&\widetilde{\mathcal{G}} (k_1^0, k_2^0, k_3^0) \equiv \frac{-1}{\beta (k_1^0 - k_2^0 - k_3^0)} + \frac{1 - e^{-\beta (k_1^0 - k_2^0 - k_3^0) }}{\beta^2 (k_1^0 - k_2^0 - k_3^0 )^2}  \, . \\[1.5ex]
\end{split}
\end{equation}
These functions are actually safe to change the integration contour in Eq.~(\ref{sumformula2}).

We can easily see the useful relations such as $\widetilde{\mathcal{G}} (-k_1^0, -k_2^0, -k_3^0) = \widetilde{\mathcal{F}} (k_1^0, k_2^0, k_3^0) $.
We write these functions as $\widetilde{\mathcal{F}}_{+++} \equiv \widetilde{\mathcal{F}} (k_1^0, k_2^0, k_3^0)$,
$\widetilde{\mathcal{F}}_{-++} \equiv \widetilde{\mathcal{F}} (-k_1^0, k_2^0, k_3^0)$,
$\widetilde{\mathcal{G}}_{-+-} \equiv \widetilde{\mathcal{G}} (-k_1^0, k_2^0, -k_3^0)$ and so on.
For convenience, we also define
\begin{equation}
\begin{split}
\mathcal{F}_{+++} \equiv 8 \left( 2 M^2 - k_1 \cdot k_2 \right) \beta \, \widetilde{\mathcal{F}}_{+++} \, , \qquad
\mathcal{G}_{+++} \equiv 8 \left( 2 M^2 - k_1 \cdot k_2 \right) \beta \, \widetilde{\mathcal{G}}_{+++} \, ,
\end{split}
\end{equation}
for QED and 
\begin{equation}
\begin{split}
\mathcal{F}_{+++} \equiv 4 \left( M^2 + k_1 \cdot k_2 \right) \beta \, \widetilde{\mathcal{F}}_{+++} \, , \qquad
\mathcal{G}_{+++} \equiv 4 \left( M^2 + k_1 \cdot k_2 \right) \beta \, \widetilde{\mathcal{G}}_{+++} \, ,
\end{split}
\end{equation}
for the Yukawa theory.
The relations such as $\mathcal{G}_{--+} = \mathcal{F}_{++-}$ are satisfied.

\section{Infrared behavior}\label{infra}

We here describe infrared behavior of the entanglement entropy in QED
and show that the equation \eqref{SAb} with \eqref{final} is finite in the case of a nonzero fermion mass $M \neq 0$.
The relevant terms for infrared divergence in \eqref{final} are
\begin{equation}
\begin{split}
\left\{ \frac{\left(  2M^2 - E_1 E_2 + \bm{k}_1 \cdot \bm{k}_2 \right)\left(E_1 - E_2 \right)
}{\left(E_1 - E_2 + \omega \right)^2}
+\frac{\left(  2M^2 - E_1 E_2 + \bm{k}_1 \cdot \bm{k}_2 \right)\left(E_1 - E_2 \right)}{\left(E_1 - E_2 - \omega \right)^2}
\right\}  n_1 N  \, . \label{IRfinite}
\end{split}
\end{equation}
To show infrared finiteness of these terms,
we change the integration variables $\bm{k}_1$, $\bm{k}_2$ into
$\bm{k} = \bm{k}_1 - \bm{k}_2$ and $\bm{p} = \frac{1}{2} (\bm{k}_1 + \bm{k}_2)$ and define
$ \bm{p} \cdot \bm{k} = p k  \cos \theta$.
Since infrared divergence can be caused by zero momentum photons, we concentrate on the region around $k=0$.
In this region, we approximately find
\begin{equation}
\begin{split}
&n_1 \simeq \frac{1}{e^{\beta E_p} + 1} \, , \qquad n_2 \simeq \frac{1}{e^{\beta E_p} + 1} \, , \qquad N \simeq \frac{1}{\beta k} \, , \\[1ex]
&E_1 \simeq E_p \left(1 + \frac{p k  \cos \theta}{2E_p^2} \right) , \qquad 
E_2 \simeq E_p \left(1 - \frac{p k  \cos \theta}{2E_p^2} \right) . \label{kzeroapp}
\end{split}
\end{equation}
Inserting these approximate expressions into \eqref{SAb} with \eqref{IRfinite}, we obtain
\begin{equation}
\begin{split}
- \frac{\beta V e^2}{4 \pi^4} &\int^\infty_0 dk \, k^2 \int^{1}_{-1} d (\cos \theta) \int^\infty_0 dp \, p^2
\, \frac{1}{E_p^2 \, k} \frac{M^2}{k} \frac{p\cos \theta}{E_p} \\[1ex]
&\times \left[ \frac{1}{\left( 1 - \frac{p\cos \theta}{E_p} \right)^2} + \frac{1}{\left( 1 + \frac{p\cos \theta}{E_p} \right)^2} \right]
\frac{1}{e^{\beta E_p} + 1} \frac{1}{\beta k} = 0 \, .
\end{split}
\end{equation}
Here, we have rewritten the first term in terms of $\cos \theta' = - \cos \theta$ to derive the equality.
Although each term in the parenthesis of the integrand is divergent around $k=0$, they are canceled with each other.
On the other hand, in the limit of $M \rightarrow 0$, infrared divergence remains.
We need to include a thermal mass in this case.


\bibliography{ref}
\bibliographystyle{utphys}

\end{document}